\documentclass[useAMS,usenatbib,usegraphicx]{mn2e}

\usepackage{dcolumn}
\usepackage{bm}
\usepackage{amssymb}
\usepackage{color}

\usepackage{graphicx}
\usepackage{amssymb}
\usepackage{amsmath}
\usepackage{times}
\usepackage{color}

\newcommand{\mnras}{MNRS}

\newcommand{\ie}{\textit{i.e.},~}
\newcommand{\eg}{\textit{e.g.},~}
\newcommand{\cf}{\textit{cf.}~}

\title[Beyond ideal MHD]{Beyond ideal MHD: towards a more
  realistic modeling of relativistic astrophysical plasmas}

\author[Palenzuela, Lehner, Reula, Rezzolla]{Carlos~Palenzuela${}^1$, Luis~Lehner${}^2$, Oscar~Reula${}^3$ and Luciano~Rezzolla${}^{1,2}$ 
\\
${}^1$Max-Planck-Institut f\" ur Gravitationsphysik,
Albert-Einstein-Institut, Golm, Germany\\
${}^2$Department of Physics and Astronomy, Louisiana State
University, Baton Rouge, LA 70803-4001, \\
${}^3$IFFaMAF, FaMAF, Universidad Nacional de Cordoba, Cordoba, 5000, Argentina 
\\}
\pagerange{\pageref{firstpage}--\pageref{lastpage}}

\pubyear{2008}

\begin{document}

\maketitle

\label{firstpage}

\begin{abstract}
    Many astrophysical processes involving magnetic fields and
    quasi-stationary processes are well described when assuming the
    fluid as a perfect conductor. For these systems, the
    ideal-magnetohydrodynamics (MHD) description captures the dynamics
    effectively and a number of well-tested techniques exist for its
    numerical solution. Yet, there are several astrophysical processes
    involving magnetic fields which are highly dynamical and for which
    resistive effects can play an important role. The numerical
    modeling of such non-ideal MHD flows is significantly more
    challenging as the resistivity is expected to change of several
    orders of magnitude across the flow and the equations are then
    either of hyperbolic-parabolic nature or hyperbolic with stiff
    terms. We here present a novel approach for the solution of these
    relativistic resistive MHD equations exploiting the properties of
    implicit-explicit (IMEX) Runge Kutta methods. By examining a
    number of tests we illustrate the accuracy of our approach under a
    variety of conditions and highlight its robustness when compared
    with alternative methods, such as the Strang-splitting. Most
    importantly, we show that our approach allows one to treat, within
    a unified framework, both those regions of the flow which are
    fluid-pressure dominated (such as in the interior of compact
    objects) and those which are instead magnetic-pressure dominated
    (such as in their magnetospheres). In view of this, the approach
    presented here could find a number of applications and serve as a
    first step towards a more realistic modeling of relativistic
    astrophysical plasmas.
 \end{abstract}


%% \date{Accepted 0000 00 00.
%%       Received 0000 00 00.}

%%%%%%%%%%%%%%%%%%%%%%%%%%%%%%%%%%%%%%%%%%%%%%%%%%%%%%%%%%%%%%%%%%%%%%%%%%%%%%%
%%%%%%%%%%%%%%%%%%%%%%%%%%%%%%%%%%%%%%%%%%%%%%%%%%%%%%%%%%%%%%%%%%%%%%%%%%%%%%%
%%%%%%%%%%%%%%%%%%%%%%%%%%%%%%%%%%%%%%%%%%%%%%%%%%%%%%%%%%%%%%%%%%%%%%%%%%%%%%%

\section{Introduction}\label{section1}

A vast number of astronomical observations suggests that magnetic
fields play a crucial role in the dynamics of many phenonema of
relativistic astrophyics, either on stellar scales, such as for
pulsars, magnetars, compact X-ray binaries, short and long/gamma-ray
bursts (GRBs) and possibly for the collapse of massive stellar cores,
but also on much larger scales, as it is the case for radio galaxies,
quasars and active galactic nuclei (AGNs). A shared aspect in all
these phenomena is that the plasma is essentially electrically neutral
and the frequency of collisions is much larger than the inverse of the
typical timescale of the system. The MHD approximation is then an
excellent description of the global properties of these plasmas and
has been employed with success over the several decades to describe
the dynamics of such systems well in their nonlinear regimes. Another
important common aspect in these systems is that their flows are
characterized by large magnetic Reynolds numbers ${\cal R}_{\rm M} = L
V/ \lambda = 4 \pi \sigma L V/c^2$, where $L$ and $V$ are the typical
sizes and velocities, respectively, while $\lambda$ is the magnetic
diffusivity and $\sigma$ is the electrical conductivity.
For a typical relativistic compact object, ${\cal R}_{\rm M} \gg 1$ and,
under these conditions, the magnetic field is
essentially advected with the flow, being continuosly distorted and
possibly amplified, but also essentially not decaying. We note that
these conditions are very different from those traditionally produced
in the Earth's laboratories, where ${\cal R}_{\rm M} \ll 1$, and the
resistive diffusion represents an important feature of the
magnetic-field evolution.

A particularly simple and yet useful limit of the MHD approximation is
that of the \textit{``ideal-MHD''} limit. This is mathematically
defined as the limit in which the electrical resistivity $\eta \equiv
1/\sigma$ vanishes or, equivalently, by an infinite electrical
conductivity. It is within this framework that many multi-dimensional
numerical codes have been developed over the last decade to study a
number of phenomena in relativistic astrophysics and in fully
nonlinear regimes \citep{Kom:1999b,KoiShiKud:1999,Kom:2001,
  KolRomUstLov:2002,GamMckTot:2003,DelBucLon:2003,AnnFraSal:2005,
  DueLiuShaSte:2005,
  ShiSek:2005,NeiHirMil:2006,AZMMIFP:2006,McKinney:2006, MigBod:2006,
  Noble:2007zx, GiaRez:2007, DelZanBucLon:2007, FarLiLiuShap:2008}.
The ideal-MHD approximation is not only a convenient way of writing
and solving the equations of relativistic MHD, but it is also an
excellent approximation for any process that takes place over a
dynamical timescale. In the case of an old and ``cold'' neutron star,
for example, the electrical and thermal transport properties of the
matter are mainly determined by the transport properties of the
electrons, which are the most important carriers of charge and
heat. At temperatures above the crystallization temperature of the
ions, the electrical (and thermal) conductivities are governed by
electron scattering off ions and an approximate expression for the
electrical conductivity is given by~\citep{L:91} $\sigma \approx
10^{24} \left({10^9\ \rm K}/{T}\right)^2 \left({\rho}/{10^{14}\ {\rm
    g\ cm}^{-3}}\right)^{3/4} \ {\rm s}^{-1}$, where $T$ and $\rho$
are the stellar temperature and mass density\footnote{Note that this
  expression for the electrical conductivity is roughly correct for
  densities in the range $10^{10}-10^{14}$ g cm$^{-3}$ and
  temperatures in the range $10^{6}-10^{8}$ K, but provides a
  reasonable estimate also at larger temperatures of $\sim
  10^{9}-10^{10}$ K [\cf.~\citet{PBHY:99}].}. Even for a magnetic
field that varies on a length-scale as small as $L \simeq 0.1 R$,
(where $R$ is the stellar radius) the magnetic diffusion timescale is
$\tau_{\rm diff} = {4\pi L^2 \sigma}/{c^2} \approx 3 \times 10^{6}
{\rm yr}$.

Clearly, at these temperatures and densities, Ohmic diffusion will be
neglible for any process taking place on a dynamical timescale for the
star, \ie $\lesssim {\rm few\ s}$, and thus the conductivity can be
considered as essentially infinite. However, 
catastrophic events, such as
the merger of two neutron stars, or of a neutron star
with a black hole, can produce plasmas with regions
at much larger temperatures (\eg $T\sim 10^{11-13}\ {\rm K}$) and much
lower densities (\eg $\rho \sim 10^{8-10}\ {\rm g cm}^{-3})$. In such
regimes, all the transport properties of the matter will be
considerably modified and non-ideal effects, absent in perfect-fluid
hydrodynamics (such as bulk viscosity) and ideal MHD (such as Ohmic
diffusion on a much shorter timescale $\tau_{\rm diff} \sim 10^3 {\rm
  s}$) will need to be taken into account. Similar conditions are
likely not limited to binary mergers but, for instance, be present also behind
processes leading to long GRBs, thus extending the range of
phenomena for which resistive effects could be important. Note also
that these non-ideal effects in hydrodynamics (MHD) are proportional
not only to the viscosity (resistivity) of the plasma, but also to the
second derivatives of the velocity (magnetic) fields. Hence, even in
the presence of a small viscosity (resistivity), their contribution to
the overall conservation of energy and momentum can be considerable if
the velocity (magnetic) fields undergo very rapid spatial variations
in the flow. A classical example of the importance of resistive MHD
effects in plasmas with high but finite conductivities is offered by
{\it current sheets}. These phenomena are often observed in the solar
activity and are responsible for the reconnection of magnetic field
lines and changes in the magnetic field topology. While these
phenomena are behind the emission of large amounts of energy, they are
strictly forbidden within the ideal-MHD limit due to magnetic flux
conservation and so can not be studied employing this limit.

Besides having considerably smaller conductivities, low-density higly
magnetized plasmas are present rather generically around magnetized
objects, constituting what is referred to as the ``magnetosphere''. 
In such regions magnetic stresses are much larger than magnetic
pressure gradients and cannot be properly balanced; as a result, the magnetic
fields have to adjust themselves so that the magnetic stresses
vanish identically. This scenario is known as the \textit{force free}
regime (because the Lorentz force vanishes in this case) and while the
equations governing it can be seen as the low-inertia limit of the
ideal-MHD equations~\citep{Kom:2002,McKinney:2006tf}, the force-free
limit is really distinct from the ideal-MHD one. This represents a
considerable complication since it implies that it is usually not
possible to decribe, within the same set of equations, both the
interior of compact objects and their magnetospheres.

Theoretical work to derive a fully relativistic theory of non-ideal
hydrodynamics and non-ideal MHD has been carried out by several
authors in the past \citep{Israel:1976,Stewart:1977,Carter:1991,Lichnerowizc:1967,Anile:1989}
and is particularly simple in the case of the resistive MHD description. The purpose of
this work is indeed that of proposing the solution of the relativistic
resistive MHD equations as an important step towards a more realistic
modelling of astrophysical plasmas. There are a number of advantages
behind such a choice. First, it allows one to use a single
mathematical framework to describe both regions where the conductivity is
large (as in the interior of compact objects) and small (as
in magnetospheres), and even the vacuum regions outside the
compact objects where the MHD equations trivially reduce to the
Maxwell equations. Second, it makes it possible to account
self-consistently for those resistive effects, such as current sheets,
which are energetically important and could provide a substantial
modification of the whole dynamics. Last but not least, the numerical
solution of the resistive MHD equations provides the only way to
control and distinguish the physical resistivity from the numerical
one. The latter, which is inevitably present and proportional to
truncation error, is also completely dependent on the specific details
of the numerical algorithm employed and on the resolution used for the
solution.

As noted already by several authors, the numerical solution of the
ideal-MHD equations is considerably less challenging than that of the
resistive MHD equations. In this latter case, in fact, the equations
become mixed hyperbolic-parabolic in Newtonian physics or hyperbolic
with stiff relaxation terms in special relativity. The presence of
stiff terms is the natural consequence of the fact that the diffusive
effects take place on timescales that are intrinsically larger than
the dynamical one. Stated differently, in such equations the
relaxation terms can dominate over the purely hyperbolic ones,
posing severe constraints on the timestep for the evolution. While
considerable work has already been made to introduce numerical
techniques to achieve efficient implementations in either
regime~\citep{Kom:2004, KomBarLyu:2007, Kom:2007, ReySamWoo:2006, GraTreMilCol:2008},
the use of these techniques in fully three-dimensional simulations is still
difficult and expensive.

In order to benefit from the many advantages discussed above in the use
of the resistive MHD equations, we here present a novel approach for
the solution of the relativistic resistive MHD equations exploiting
the properties of implicit-explicit (IMEX) Runge Kutta methods. This
approach represents a simple but effective solution to the problem of
the vastly different timescales without sacrificing either
simplicity in the implementation or the numerical efficiency. By
examining a number of tests we illustrate the accuracy of our approach
under a variety of conditions and demonstrate its robustness. In
addition, we also compare it with the alternative method proposed
by~\citet{Kom:2007} for the solution of the same set of relativistic
resistive MHD equations. This latter approach employs
Strang-splitting techniques and the analytical integration of a reduced
form of Ampere's law. While it works well
in a number of cases, it has revealed to be unstable when
applied to discontinuous flows with large conductivities; such
difficulties were not encountered when solving the same problem within the
IMEX implementation.

Because our approach effectively treats within a unified framework
both those regions of the flow which are fluid-pressure dominated and
those which are instead magnetic-pressure dominated, it could find a
number of applications and serve as a first step towards a more
realistic modeling of relativistic astrophysical plasmas.

Our work is organized as follows. In Sect.~\ref{section2} we present
the system of equations describing a resistive magnetized fluid, while
in Section \ref{section3} we discuss the problems related to the
numerical evolution of this system of equations and the numerical
approaches developed to solve them. In particular, we introduce
the basic features of the IMEX Runge-Kutta schemes and recall their
stability properties. In Sect~\ref{section4} we instead explain in
detail the implementation of the IMEX scheme to the resistive MHD
equations. Finally, in Sect.~\ref{section5} we present the numerical
tests carried out either in one or two dimensions and that span
several prescriptions for the conductivity. Section \ref{section5} is
also dedicated to the comparison with the Strang-splitting technique.
The conclusions and the perspectives for future improvements are
presented in Sect.~\ref{section6}, while Appendix~\ref{appendixB}
reviews our space discretization of the equations.

Hereafter we will adopt Gaussian units such that $c=1$ and employ the
summation convention on repeated indices. Roman indices $a,b,c,...$
are used to denote spacetime components (\ie from $0$ to $3$), while
$i,j,k,...$ are used to denote spatial ones; lastly, bold italics
letters represent vectors, while bold letters represent tensors.

%%%%%%%%%%%%%%%%%%%%%%%%%%%%%%%%%%%%%%%%%%%%%%%%%%%%%%%%%%%%%%%%%%%%%%%%%%%%%%%
%%%%%%%%%%%%%%%%%%%%%%%%%%%%%%%%%%%%%%%%%%%%%%%%%%%%%%%%%%%%%%%%%%%%%%%%%%%%%%%
%%%%%%%%%%%%%%%%%%%%%%%%%%%%%%%%%%%%%%%%%%%%%%%%%%%%%%%%%%%%%%%%%%%%%%%%%%%%%%%

\section{The resistive MHD description}\label{section2}

An effective description of a fluid in the presence of electromagnetic
fields can be made by considering three different sets of equations
governing respectively the electromagnetic fields, the fluid variables
and the coupling between the two. In particular, the electromagnetic
part can be described via the Maxwell equations, while the
conservation of energy and momentum can be used to express the
evolution of the fluid variables. Finally, Ohm's law, whose exact form
depends on the microscopic properties of the fluid, expresses the
coupling between the electromagnetic fields and the fluid
variables. In what follows we review these three sets of equations
separately, discuss how they then lead to the resistive MHD
description, and how the latter reduces to the well-known limits of
ideal-MHD and of the Maxwell equations in vacuum. Our presentation
will be focussed on the special-relativistic regime, but the extension
to general relativity is rather straightforward and will be presented
elsewhere.

%%%%%%%%%%%%%%%%%%%%%%%%%%%%%%%%%%%%%%%%%%%%%%%
\subsection*{The Maxwell equations}

The special relativistic Maxwell equations can be written
as~\citep{LanLif:1980}
\begin{eqnarray}\label{maxwell_covariant1}
   \partial_b F^{ab} &=& I^a\,, \\
\label{maxwell_covariant2}
   \partial_b ^{~*}\!F^{ab} &=& 0\,, 
\end{eqnarray}
where $F^{ab}$ and $^{~*}\!F^{ab}$ are the Maxwell and the Faraday
tensor respectively and $I^a$ is the electric current 4-vector. A
highly-ionized plasma has essentially zero electric and magnetic
susceptibilities and the Faraday tensor is then simply the dual of the
Maxwell tensor. This tensor provides information about the electric
and magnetic fields measured by an observer moving along any timelike
vector $n^a$, namely
\begin{eqnarray}\label{maxwell_tensor} 
   F^{ab} = n^a E^b - n^b E^a + \epsilon^{abc} B_c\,.
\end{eqnarray}
We are considering  $n^a$ to be the time-like traslational
killing vector field in a flat (Minkowski) spacetime, so $n_a=(-1,0,0,0)$
and the Levi-Civita symbol $\epsilon^{abc}$ is non-zero only for spatial indices.
Note that the electromagnetic fields have no components
parallel to $n^a$ (\ie $E^a~n_a=0=B^a~n_a$).

By using the decomposition of the Maxwell tensor
(\ref{maxwell_tensor}), the equations
(\ref{maxwell_covariant1})--(\ref{maxwell_covariant2}) can be split
into directions which are parallel and orthogonal to $n^a$ to yield
the familiar Maxwell equations
\begin{eqnarray}
   \nabla \cdot {\boldsymbol E} &=& q\,,
\label{maxwell_clasic1} \\    
   \nabla \cdot {\boldsymbol B} &=& 0\,,
\label{maxwell_clasic2} \\
   \partial_t {\boldsymbol E} - \nabla \times {\boldsymbol B}  &=& - {\boldsymbol J} \, ,
\label{maxwell_clasic3} \\
   \partial_t {\boldsymbol B} + \nabla \times {\boldsymbol E} &=& 0 \,,
\label{maxwell_clasic4} 
\end{eqnarray}
where we have decomposed also the current vector $I^a = q n^a + J^a$, with
$q$ being the charge density, $qn^a$ the convective current and $J^a$
the conduction current satisfying $J^a~n_a=0$.

The current conservation equation $\partial_a I^a = 0$ follows from
the antisymmetry of the Maxwell tensor and provides the evolution of
the charge density $q$
\begin{eqnarray}\label{current_conservation_clasic}
   \partial_t q + \nabla \cdot {\boldsymbol J} = 0\,,
\end{eqnarray}
which can be obtained also directly by taking the divergence of
(\ref{maxwell_clasic3}) when the constraints
(\ref{maxwell_clasic1})--(\ref{maxwell_clasic2}) are satisfied.

%If this is not the case, as it may happen in a numerical solution,
%current conservation is not guaranteed and
%Eq.~(\ref{current_conservation_clasic}) should be regarded as an
%additional evolution equation for the electric charge.

%%%%%%%%%%%%%%%%%%%%%%%%%%%%%%%%%%%%%%%%%%%%%%%%%%%%%
\subsection*{The hydrodynamic equations}

The evolution of the matter follows from the conservation of the stress-energy
tensor
\begin{eqnarray}\label{conservation_Tmunu} 
   \partial_b  T^{ab} = 0 \, ,
\end{eqnarray}
and the conservation of baryon number
\begin{eqnarray}\label{conservation_baryons} 
   \partial_a  (\rho u^a) = 0 \, ,
\end{eqnarray}
where $\rho$ is the rest-mass density (as measured in the rest frame
of the fluid) and $u^a$ is the fluid 4-velocity. The stress-energy
tensor $T^{ab}$ describing a perfect fluid minimally coupled to an
electromagnetic field is given by the superposition
\begin{eqnarray}\label{full_Tmunu} 
   T_{ab} &=& T_{ab}^{\rm fluid} + T_{ab}^{\rm em} \, , 
\end{eqnarray}
where
\begin{eqnarray}
\label{Tmunu_em} 
    T^{ab}_{\rm em} &\equiv& F^{ac} F^b_c - \frac{1}{4} (F^{cd} F_{cd}) g^{ab} \, , \\
\label{Tmunu_fluid} 
    T^{ab}_{\rm fluid} &\equiv& h u^a u^b + p~g^{ab} \, .
\end{eqnarray}
Here $h \equiv \rho (1 + \epsilon) + p$ is the enthalpy, with $p$ the
pressure and $\epsilon$ the specific internal energy.

The conservation law (\ref{conservation_Tmunu}) can be split into
directions parallel and orthogonal to $n^a$ to yield the familiar
energy and momentum conservation laws
\begin{eqnarray}
    \partial_t \tau + \nabla \cdot \boldsymbol{F}_{\tau} = 0 \, , 
\label{fluid_tau} \\
    \partial_t {\boldsymbol S} + \nabla \cdot {\bf F}_{\boldsymbol S} = 0 \, ,
\label{fluid_S} 
\end{eqnarray}
where we have introduced the conserved quantities $\{ \tau,
{\boldsymbol S} \}$, which are essentially the energy density $ \tau
\equiv T_{ab} n^a n^b $and the energy flux density $S_i \equiv T_{ai}
n^a$, and whose expressions are given by
\begin{eqnarray}
    \tau &\equiv& \frac{1}{2} (E^2 + B^2) + h~W^2 - p \, ,
\label{def_e} \\
    {\boldsymbol S} &\equiv& {\boldsymbol E} \times {\boldsymbol B} + h~W^2~{\boldsymbol v} ~.
\label{def_S} 
\end{eqnarray}
Here ${\boldsymbol v}$ is the velocity measured by the inertial
observer and $W \equiv -n_a u^a = 1/ \sqrt{1-v^2}$ is the Lorentz
factor. The fluxes can then be written as
\begin{eqnarray}
    && \hskip -0.5cm
\boldsymbol{F}_{\tau} \equiv {\boldsymbol E} \times {\boldsymbol B} + h~W^2~{\boldsymbol v} \, ,
\label{def_Fe} \\
    && \hskip -0.5cm
{\bf F}_{\boldsymbol S} \equiv -\boldsymbol{E E} - \boldsymbol{B B} + h W^2 \boldsymbol{v v}  
+ \left[ \frac{1}{2} (E^2 + B^2) + p \right] {\bf g} \, .
\label{def_FS} 
\end{eqnarray}
Finally, the conservation of the baryon number
(\ref{conservation_baryons}) reduces to the continuity equation
written as
\begin{eqnarray}\label{fluid_baryons} 
    \partial_t D + \nabla \cdot \boldsymbol{F}_{D} = 0 \, ,
\end{eqnarray}
where we have introduced another conserved quantity $D \equiv  \rho W$ and its
flux $\boldsymbol{F}_{D} \equiv  \rho W {\boldsymbol v}$.

%%%%%%%%%%%%%%%%%%%%%%%%%%%%%%%%%%%%%%%%%%%%%%%
\subsection*{Ohm's law}

As mentioned above, Maxwell equations are coupled to the fluid ones by
means of the current 4-vector ${I}^a$, whose explicit form will depend
in general on the electromagnetic fields and on the local fluid
properties. A standard prescription is to consider the current to be
proportional to the Lorentz force acting on a charged particle and the
electrical resistivity $\eta$ to be a scalar function. Ohm's law,
written in a Lorentz invariant way, then reads
\begin{equation}\label{ohm_relativistic_covariant}
 I_a + (I^b~u_b) u_a=  \sigma~F_{ab}~u^b \, ,
\end{equation}
with $\sigma \equiv 1/\eta$ being the electrical conductivity of the medium.
Expressing (\ref{ohm_relativistic_covariant}) in terms of the electric
and magnetic fields one obtains the familiar form of Ohm's law in a general inertial frame  
\begin{equation}\label{ohm_relativistic}
 {\boldsymbol J} = \sigma~W [{\boldsymbol E} + {\boldsymbol v} \times {\boldsymbol B}
         - ({\boldsymbol E} \cdot {\boldsymbol v}) {\boldsymbol v}] + q~{\boldsymbol v} \, .
\end{equation}
Note that the conservation of the electric charge
(\ref{current_conservation_clasic}) provides the evolution equation
for the charge density $q$ (\ie the projection of the 4-current
$\boldsymbol{I}$ along the direction ${\boldsymbol n}$), while Ohm's
law provides a prescription for the (spatial) conduction current
${\boldsymbol J}$ (\ie the components of ${\boldsymbol I}$ orthogonal
to ${\boldsymbol n}$).

It is important to recall that in deriving
expression~(\ref{ohm_relativistic}) for Ohm's law we are implicitly
assuming that the collision frequency of the constituent particles of
our fluid is much larger that the typical oscillation frequency of the
plasma. Stated differently, the timescale for the electrons and ions
to come into equilibrium is much shorter than any other timescale in
the problem, so that no charge separation is possible and the fluid is
globally neutral. This assumption is a key aspect of the MHD
approximation.

The well-known ideal-MHD limit of Ohm's law can be obtained by
requiring the current to be finite even in the limit of infinite
conductivity ($\sigma \rightarrow \infty $).  In this limit Ohm's law
(\ref{ohm_relativistic}) then reduces to
\begin{equation}
  {\boldsymbol E} + {\boldsymbol v} \times {\boldsymbol B} - ({\boldsymbol E} \cdot {\boldsymbol v}) {\boldsymbol v} = 0 \, .
\end{equation}
Projecting this equation along ${\boldsymbol v}$ one finds that the
electric field does not have a component along that direction and then from
the rest of the equation one recovers the well-known ideal-MHD
condition
\begin{equation}
\label{ef_imhd}
 {\boldsymbol E} = - {\boldsymbol v} \times {\boldsymbol B} \,,
\end{equation}
stating that in this limit the electric field is orthogonal to both
${\boldsymbol B}$ and ${\boldsymbol v}$. Such a condition also
expresses the fact that in ideal MHD the electric field is not an
independent variable since it can be be computed via a simple
algebraic relation from the velocity and magnetic vector fields.

Summarizing: the system of equations of the relativistic resistive MHD
approximation is given by the constraint equations
(\ref{maxwell_clasic1})--(\ref{maxwell_clasic2}),
 evolution equations
(\ref{maxwell_clasic3})--(\ref{current_conservation_clasic}),
(\ref{fluid_tau})--(\ref{fluid_S}) and (\ref{fluid_baryons}), where
the fluxes are given by Eqs.~(\ref{def_Fe})--(\ref{def_FS}) and the
3-current is given by Ohm's law (\ref{ohm_relativistic}). These
equations, together with a equation of state (EOS) for the fluid and
a reasonable model for the conductivity, completely describe the system
under consideration provided consistent initial and boundary data are defined.

%%%%%%%%%%%%%%%%%%%%%%%%%%%%%%%%%%%%%%%%%%%%%%%
\subsection*{Different limits of the resistive MHD description}

At this point it is useful to point out some properties of the
relativistic resistive MHD equations discussed so far, to underline
their purely hyperbolic character and to contrast them with those of
other forms of the resistive MHD equations which contain a parabolic
part instead. To do this within a simple example, we adopt the
Newtonian limit of Ohm's law (\ref{ohm_relativistic}),
\begin{equation}\label{ohm_classic}
 {\boldsymbol J} = \sigma [{\boldsymbol E} + {\boldsymbol v} \times {\boldsymbol B}] \, ,
\end{equation}
where we have neglected terms of order ${\cal O}(v^2/c^2)$, obtaining 
the following potentially stiff equation for the electric field
\begin{equation}
   \partial_t {\boldsymbol E} - \nabla \times {\boldsymbol B}  =
    - \sigma [{\boldsymbol E} + {\boldsymbol v} \times {\boldsymbol B}] \, .
\label{maxwell_stiff} 
\end{equation}
Assuming now a uniform conductivity and taking a time derivative of
Eq.~(\ref{maxwell_clasic4}), we obtain the following hyperbolic
equation with relaxation terms (henceforth referred simply 
as hyperbolic-relaxation equation) for the magnetic field
\begin{equation}\label{relativistic_rMHD} \\
 -\frac{1}{\sigma} [\partial_{tt} {\boldsymbol B} - \nabla^2 {\boldsymbol B}]
  = [\partial_t {\boldsymbol B}  - \nabla \times ({\boldsymbol v} \times {\boldsymbol B})]\,.
\end{equation}

If the displacement current can be neglected, \ie $\partial_t {\boldsymbol E}
\simeq \partial_{tt} {\boldsymbol B} \simeq 0$, equation
(\ref{relativistic_rMHD}) reduces to the familiar
parabolic equation for the magnetic field 
\begin{equation}\label{newtonian_rMHD} \\
   \partial_t {\boldsymbol B} 
    - \nabla \times ({\boldsymbol v} \times {\boldsymbol B})
    - \frac{1}{\sigma} \nabla^2 {\boldsymbol B} = 0  \,,
\end{equation}
where the last term is responsible for the diffusion of the magnetic
field. It is important to stress the significant difference in the
characteristic structure between equations (\ref{relativistic_rMHD})
and (\ref{newtonian_rMHD}). Both equations reduce to the same
advection equation in the ideal-MHD limit of infinite conductivity
($\sigma \rightarrow \infty$) indicating the flux-freezing
condition. However, in the opposite limit of infinite resistivity
($\sigma \rightarrow 0$) Eq.~(\ref{newtonian_rMHD}) tends to the
(physically incorrect) elliptic Laplace equation $\nabla^2
{\boldsymbol B} = 0$ while Eq.~(\ref{relativistic_rMHD}) reduces to
the (physically correct) hyperbolic wave equation for the magnetic
field.

%%%%%%%%%%%%%%%%%%%%%%%%%%%%%%%%%%%%%%%%%%%%%%%
%%%%%%%%%%%%%%%%%%%%%%%%%%%%%%%%%%%%%%%%%%%%%%%
%%%%%%%%%%%%%%%%%%%%%%%%%%%%%%%%%%%%%%%%%%%%%%%
\subsection{The augmented MHD system}

The set of Maxwell equations described above can also be cast in an
extended fashion which includes two additional fields, $\psi$ and
$\phi$, introduced to control dynamically the constraints of the
system, \ie Eqs~ (\ref{maxwell_clasic1}) and (\ref{maxwell_clasic2}). This
\textit{``augmented''} system reads
\begin{eqnarray}\label{maxwell_augmented1}
   \partial_b (F^{ab} + \psi g^{ab}) &=& I^a - \kappa \psi n^a\,, \\
\label{maxwell_augmented2}
   \partial_b (^{~*}\!F^{ab} + \phi g^{ab}) &=& - \kappa \phi n^a\,.
\end{eqnarray}
Clearly, the standard Maxwell equations
(\ref{maxwell_covariant1})--(\ref{maxwell_covariant2}) are recovered
when $\psi = \phi = 0$ and we are in this way extending the space of
solutions of the original Maxwell equations to include those with
non-vanishing $\{\psi,\phi\}$.

The evolution of these extra scalar fields can be obtained by taking a
partial derivative $\partial_a$ of the augmented Maxwell equations
(\ref{maxwell_augmented1})--(\ref{maxwell_augmented2}) and using the
antisymmetry of the Maxwell and Faraday tensors together with the conservation
of charge to obtain
\begin{eqnarray}
   \partial_a \partial^a \psi &=& - \kappa \partial_a (\psi n^a)\,, \\
   \partial_a \partial^a \phi &=& - \kappa \partial_a (\phi n^a)\,.
\end{eqnarray}
It is evident that these represent wave equations with sources for the
scalar fields $\{\psi,\phi\}$, which propagate at the speed of light
while being damped if $\kappa > 0$. In particular, for any positive
$\kappa$, they decay exponentially over a timescale $\sim 1/\kappa$ to
the trivial solution $\psi = \phi = 0$ and the augmented system then
reduces to the standard Maxwell equations, including the constraints
(\ref{maxwell_clasic1}) and (\ref{maxwell_clasic2}). This approach,
named hyperbolic divergence cleaning in the context of ideal
MHD~\citep{DKKMSW:2002}, was proposed as a simple way of solving
the Maxwell equations and enforcing the conservation of the
divergence-free condition for the magnetic field.

Adopting this approach and following the formulation proposed
by~\citet{Kom:2007}, the evolution equations of the augmented Maxwell
equations (\ref{maxwell_augmented1})--(\ref{maxwell_augmented2}) can
then be written as
\begin{eqnarray}
   \partial_t \psi + \nabla \cdot {\boldsymbol E}  &=& q - \kappa~\psi \, ,
\label{maxwell_aug1} \\    
   \partial_t \phi + \nabla \cdot {\boldsymbol B} &=& -\kappa~\phi \, ,
\label{maxwell_aug2} \\
   \partial_t {\boldsymbol E} - \nabla \times {\boldsymbol B} + \nabla \psi &=& - {\boldsymbol J} \, ,
\label{maxwell_aug3} \\
   \partial_t {\boldsymbol B} + \nabla \times {\boldsymbol E} + \nabla \phi &=& 0 \, .
\label{maxwell_aug4} 
\end{eqnarray}
The system of equations (\ref{maxwell_aug1})--(\ref{maxwell_aug4}), together
with the current conservation (\ref{current_conservation_clasic}), is
the one we will use for the numerical evolution of the electromagnetic
fields within the set of relativistic resistive MHD equations.

%%%%%%%%%%%%%%%%%%%%%%%%%%%%%%%%%%%%%%%%%%%%%%%%%%%%%%%%%%%%%%%%%%%%%%%%%%%%%%%
%%%%%%%%%%%%%%%%%%%%%%%%%%%%%%%%%%%%%%%%%%%%%%%%%%%%%%%%%%%%%%%%%%%%%%%%%%%%%%%

\section{Evolution of hyperbolic-relaxation equations}\label{section3}

While the ideal-MHD equations are well suited to an efficient
numerical implementation, the general system of relativistic resistive
MHD equations brings about a delicate issue when the conductivity in
the plasma undergoes very large spatial variations. In the regions
with high conductivity, in fact, the system will evolve on timescales
which are very different from those in the low-conductivity
region. Mathematically, therefore, the problem can be regarded as a
hyperbolic one with stiff relaxation terms which requires special
care to capture the dynamics in a stable and accurate manner. In the
next Section we discuss a simple example of a hyperbolic equation with
relaxation which exhibits the problems discussed above and then
introduce implicit-explicit (IMEX) Runge Kutta methods to deal with
these kind of equations. In essence, these methods treat the advection
character of the system with strong-stability preserving (SSP)
explicit schemes, while the relaxation character with an L-stable
diagonally implicit Runge Kutta (DIRK)
scheme. After presenting the scheme, its properties and some examples,
we discuss in detail its application to the resistive MHD equations.

%%%%%%%%%%%%%%%%%%%%%%%%%%%%%%%%%%%%%%%%%%%%%%%%%%%%%%%%%%%%%%%%%%%%%%%%%%%%%%%

%%%%%%%%%%%%%%%%%%%%%%%%%%%%%%%%%%%%%%%%%%%%%%%
\subsection{Hyperbolic systems with relaxation terms}

A prototypical hyperbolic equation with relaxation is given by 
\begin{eqnarray}\label{stiff_equation}
    \partial_t {\boldsymbol U} = F({\boldsymbol U}) + \frac{1}{\epsilon} R({\boldsymbol U})\,,
\end{eqnarray}
where $\epsilon >0$ is the \textit{relaxation time} (not necessarily
constant either in space or in time), $F({\boldsymbol U})$ gives
rise to a quasilinear system of equations (\ie $F({\boldsymbol U})$
depends linearly on first derivatives of ${\boldsymbol U}$), and $R$
does not contain derivatives of ${\boldsymbol U}$.

In the limit $\epsilon \rightarrow \infty$ (corresponding for the
resistive MHD equations to the case of vanishing conductivity) the
system is hyperbolic with propagation speeds bounded by $c_h$. This
maximum bound, together with the length scale $L$ of the system,
define a characteristic timescale $\tau_h \equiv L / c_h$ of the
hyperbolic part. In the opposite limit $\epsilon \rightarrow 0$
(corresponding to the case of infinite conductivity), the system is
instead said to be \textit{stiff}, since the timescale $\epsilon$ of
the relaxation (or stiff) term $R({\boldsymbol U})$ is in general much
larger than the timescale $\tau_h$ of the hyperbolic part
$F({\boldsymbol U})$. In such a limit, the stability of an explicit
scheme is only achieved
\footnote{Implicit schemes could avoid this issue at an increased
computational cost; however, an explicit second order accurate 
method approaching iteratively the Crank-Nicholson scheme
has been shown, in a simple model with hyperbolic-relaxation terms, 
to work well when dealing with smooth profiles without being too costly 
(M. Choptuik, private communication)}
with a timestep size $\Delta t \leq \epsilon$.
This requirement is certainly more restrictive than the
Courant-Lewy-Friedrichs (CFL) stability condition $\Delta t \leq
\Delta x / c_h $ for the hyperbolic part and makes an explicit
integration impractical.  The development of efficient numerical
schemes for such systems is challenging, since in many applications
the relaxation time can vary by several orders of magnitude across the
computational domain and, more importantly, to much beyond the one 
determined by the speed $c_h$.

When faced with this issue several strategies can be adopted.  The
most straightforward one is to consider only the stiff limit $\epsilon
\rightarrow 0$, where the system is well approximated by a suitable
reduced set of conservation laws called \textit{``equilibrium system''}
\citep{CheLevLiu:1994} such that
\begin{eqnarray}\label{equilibrium_system}
     R(\boldsymbol{\bar{U}}) &=& 0 \,,\\
    \partial_t \boldsymbol{\bar{U}} &=& G(\boldsymbol{\bar{U}}) \,.
\end{eqnarray}
where $\boldsymbol{\bar{U}}$ is a reduced set of variables. This
approach can be followed if the resulting system is also hyperbolic.
This is precisely the case in the resistive MHD equations for
vanishing resistivity $\eta \rightarrow 0$ (or $\sigma \rightarrow
\infty$). In this case, the equations reduce to those of
ideal MHD and describe indeed an ``equilibrium system'' in which the magnetic
field is simply advected with the flow. As discussed earlier, this
limit is often adequate to describe the behaviour of dense
astrophysical plasmas, but it may also stray away in the
magnetospheres. A more general approach could consist of dividing the
computational domain in regions in each of which a simplified set of
equations can be adopted. As an example, the ideal-MHD equations could
be solved in the interior of compact objects, the force-free MHD
equations could be solved in the magnetosphere, and finally the
Maxwell equations for the vacuum regions outside the compact
object. However, this approach requires the overall scheme to suitably
match the different regions so as to obtain a global solution. This
task, unfortunately, is far from being straightforward and, to date,
it lacks a rigorous definition.

An alternative approach consists of considering the original 
hyperbolic-relaxation
system in the whole computational domain and then employ suitable
numerical schemes that work for all regions. Among such schemes is the
Strang-splitting technique~\citep{Strang:1968}, which has been
recently applied by~\citet{Kom:2007} for the solution of the (special)
relativistic resistive MHD equations. The Strang-splitting scheme
provides second-order accuracy if each step is at least second-order
accurate, and this property is maintained under suitable assumptions
even for stiff problems~\citep{JahLub:2000}. In practice, however,
higher-order accuracy is difficult to obtain even in non-stiff regimes
with this kind of splitting. Moreover, when applied to hyperbolic
systems with relaxation, Strang-splitting schemes reduce to
first-order accuracy since the kernel of the relaxation operator is
non-trivial and corresponds to a singular matrix in the linear case,
therefore invalidating the assumptions made by~\citet{JahLub:2000} to
ensure high-order accuracy.~\citet{Kom:2007} avoided this problem by
solving analytically the stiff part in a reduced form of Ampere's law.
Although this procedure works well for smooth solutions, our
implementation of the method has revealed problems when evolving
discontinuous flows (shocks) for large-conductivities
plasmas. Moreover, it is unclear whether the same procedure can be
adopted in more general configurations, where an analytical solution
may not be available.

As an alternative approach to the methods solving the relativistic
resistive MHD equations on a single computational domain, we here
introduce an IMEX Runge-Kutta method
\citep{AshRuuWet:1995,AshRuuSpi:1997,Par:2001,ParRus:2005} to cope
with the stiffness problems discussed above. These methods, which are
easily implemented, are still under development and have few
(relatively minor) drawbacks. The most serious one is a degradation to
first or second-order accuracy for a range of values of the relaxation
time $\epsilon$. However, since High-Resolution Shock-Capturing (HRSC) schemes
usually employed for the solution of the hydrodynamic equations already suffer from similar
effects at discontinuities, the possible degradation of the IMEX
schemes does not spoil the overall quality numerical solution when
employed in conjunction with HRSC schemes. The next sections
review in some detail the IMEX schemes and our specific implementation
for the relativistic resistive MHD equations.

%%%%%%%%%%%%%%%%%%%%%%%%%%%%%%%%%%%%%%%%%%%%%%%
\subsection{The IMEX Runge-Kutta methods}

The IMEX Runge-Kutta schemes rely on the application of an implicit
discretization scheme to the stiff terms and of an explicit one to the
non-stiff ones. When applied to system (\ref{stiff_equation}) it takes
the form \citep{ParRus:2005}
\begin{eqnarray}\label{IMEX}
&& \hskip -0.5cm
   {\boldsymbol U}^{(i)} = {\boldsymbol U}^n + \Delta t \sum_{j=1}^{i-1} {\tilde{a}}_{ij} F({\boldsymbol U}^{(j)}) 
     + \Delta t  \sum_{j=1}^{\nu} a_{ij} \frac{1}{\epsilon}
     R({\boldsymbol U}^{(j)})\,, \nonumber \\
 && \hskip -0.5cm
{\boldsymbol U}^{n+1} = {\boldsymbol U}^n + \Delta t \sum_{i=1}^{\nu} {\tilde{\omega}}_{i} F({\boldsymbol U}^{(i)})
     + \Delta t  \sum_{i=1}^{\nu} \omega_{i} \frac{1}{\epsilon} R({\boldsymbol U}^{(i)})\,, 
\nonumber\\
\end{eqnarray}
where ${\boldsymbol U}^{(i)}$ are the auxiliary intermediate values of
the Runge-Kutta scheme.  The matrices $\tilde{A}= (\tilde{a}_{ij})$
and $A= (a_{ij})$ are $\nu \times \nu$ matrices such that the
resulting scheme is explicit in $F$ (\ie $\tilde{a}_{ij} = 0$ for $j
\geq i$) and implicit in $R$. An IMEX Runge-Kutta scheme is
characterized by these two matrices and the coefficient vectors
$\tilde{\omega}_i$ and $\omega_i$.  Since simplicity and efficiency in
solving the implicit part at each step is important, it is
natural to consider diagonally implicit Runge-Kutta (DIRK) schemes
(\ie $a_{ij}=0$ for $j > i$) for the stiff terms. 

A particularly convenient way of describing an IMEX Runge-Kutta scheme
is offered by the Butcher notation, in which the scheme is by a double
tableau of the type~\citep{But:1987,But:2003}
\begin{equation}
\begin{minipage}{1.2in}
\begin{tabular} {c c c}
${\tilde c}$  & \vline & ${\tilde A}$  \\
\hline 
              & \vline & ${\tilde \omega}^T$  
\end{tabular}
\end{minipage} 
\hskip 1.0cm
\begin{minipage}{1.2in}
\begin{tabular} {c c c}
${c}$  &  \vline & ${A}$  \\
\hline 
       &  \vline & ${\omega}^T$  
\label{butcher_tableau}
\end{tabular}
\end{minipage}
\end{equation}
where the index $T$ indicates a transpose and where the coefficients
$\tilde{c}$ and $c$ used for the treatment of non-autonomous systems
are given by 
\begin{equation}
\label{definition_cs}
   {\tilde c}_{i} = \sum_{j=1}^{i-1}~ {\tilde{a}}_{ij}\,, \hskip 2.0cm
   {c}_{i} = \sum_{j=1}^{i}~ {a}_{ij} ~~~.
\end{equation}
The accuracy of each of the Runge-Kutta is achieved by imposing 
restrictions in some of the coefficients of their respective
Butcher tableaus. Although each of them separately can have an arbitrary
accuracy, this does not ensure that the combination of the two schemes
will preserve the same accuracy. In addition to the above conditions for
each Runge-Kutta scheme, there are also some additional conditions combining terms
in the two tableaus which must be fulfilled in order to achieve a global
accuracy order for the complete IMEX scheme.

Since the details of these methods are not widely known, we first
consider a simple example to fix ideas. A second-order IMEX scheme can
be written in the tableau form given in Table \ref{SSP2-222}. The
intermediate and final steps of this IMEX Runge-Kutta scheme would
then be written explicitly as
\begin{eqnarray}\label{IMEX-example}
&&
{\boldsymbol U}^{(1)} = {\boldsymbol U}^n + \frac{\Delta t}{\epsilon} \gamma R({\boldsymbol U}^{(1)}) \,,
\nonumber \\
&&
{\boldsymbol U}^{(2)} = {\boldsymbol U}^n + \Delta t F({\boldsymbol U}^{(1)}) 
\nonumber \\
 && \hskip 1.6cm  + \frac{\Delta t}{\epsilon} [(1 - 2 \gamma) R({\boldsymbol U}^{(1)}) + \gamma R({\boldsymbol U}^{(2)})] \,,
\nonumber \\ 
  &&{\boldsymbol U}^{n+1} = {\boldsymbol U}^n + \frac{\Delta t}{2} [ F({\boldsymbol U}^{(1)}) + F({\boldsymbol U}^{(2)}) ]
\nonumber \\
  && \hskip 1.8cm   + \frac{\Delta t}{2 \epsilon} [R({\boldsymbol U}^{(1)}) + R({\boldsymbol U}^{(2)})] \,.
\nonumber
\end{eqnarray}
Note that at each sub-step an implicit equation for the auxiliary
intermediate values ${\boldsymbol U}^{(i)}$ must be solved.  The
complexity of inverting this equation will clearly depend on the
particular form of the operator $R({\boldsymbol U})$.

%%%%%%%%%%%%%%%%%%%%%%%%%%%%%%%%%%%%%%%%%%%%%%%
\subsubsection{Stability properties of the IMEX schemes}

Stable solutions of conservation-type equations are usually analyzed
in terms of a suitable norm being bounded in time.  With ${\boldsymbol
  U}^n$ representing the solution vector at the time $t= n~\Delta t$,
then a sequence $\{{\boldsymbol U}^n\}$ is said to be
\textit{``strongly stable''} in a given norm $\| \cdot \|$ provided
that $\| {\boldsymbol U}^{n+1} \| \leq \| {\boldsymbol U}^n \|$ for
all $n \geq 0$.

\begin{table}
\caption{Tableau for the explicit (left) implicit (right) IMEX-SSP2 $(2,2,2)$
L-stable scheme}
\begin{minipage}{1.2in}
\begin{tabular} {c c c c}
 $0$ & \vline & $0$ & $0$  \\
 $1$ & \vline & $1$ & $0$  \\
\hline 
   & \vline & $1/2$ & $1/2$  \\
\end{tabular}
\end{minipage}
\begin{minipage}{1.2in}
\begin{tabular} {c c c c}
 $\gamma$ & \vline & $\gamma$ & $0$  \\
 $1 - \gamma$ & \vline & $1 - 2 \gamma$ & $\gamma$  \\
\hline 
   & \vline & $1/2$ & $1/2$  \\
\end{tabular}
\end{minipage}
\begin{eqnarray}
\gamma \equiv 1 - \frac{1}{\sqrt{2}}\,. \nonumber
\end{eqnarray}
\label{SSP2-222}
\end{table}

The most commonly used norms for analyzing schemes for nonlinear
systems are the Total-Variation (TV) norm and the infinity norm.  A
numerical scheme that maintains strong stability at the discrete level is
called Strong Stability Preserving (SSP) (see \citet{SpiRuu:2002} for
a detailed description of optimal SSP schemes and their properties).
Because of the stability properties of the IMEX
schemes~\citep{ParRus:2005}, it follows that if the explicit part of
the IMEX scheme is SSP, then the method is SSP for the equilibrium
system in the stiff limit. This property is essential to avoid
spurious oscillations during the evolution of non-smooth data.

The stability of the implicit part of the IMEX scheme is ensured
by requiring that the Runge-Kutta is ``L-stable'' and  this represents an
essential condition
for stiff problems. In practice, this amounts to requiring that the numerical approximation
is bounded in cases when the exact solution is bounded.
A more strict definition can be derived starting from a linear scalar
ordinary differential equation,
namely
\begin{equation}
\label{ode}
    {\rm d}_t \Psi = q \Psi \,.
\end{equation}
In this case it is easy to define the stability (or
amplification) function $C(z)$ as the ratio of the solutions 
at subsequent timesteps 
$C(z)\equiv\Psi^{n+1}/\Psi^{n}$, where $z \equiv \Delta t\,q$.
A Runge-Kutta scheme is then said to be \textit{L-stable} if
$|C(z)|<1$ (\ie it is bounded) and $C(\infty)=0$
\citep{But:1987,But:2003}.

There are a number of IMEX Runge-Kutta schemes available in the
literature and we report here only some of the second and third-order
schemes which satisfy the condition that in the limit $\epsilon
\rightarrow 0$, the solution corresponds to that of the equilibrium
system (\ref{equilibrium_system})~\citep{ParRus:2005}. These are
given in their Butcher tableau form in Table~\ref{SSP2-322} and are
taken from~\citet{ParRus:2005}.  In all these schemes the implicit
tableau corresponds to an L-stable scheme.
The tableaus are reported in the notation SSP$k(s,\sigma,p)$, where $k$
denotes the order of the SSP scheme and the triplet $(s,\sigma,p)$ characterizes
respectively the number of stages of the implicit scheme ($s$), the number of stages of
the explicit scheme ($\sigma$), and the order of the IMEX scheme
($p$).

\begin{table}
\caption{Tableaux for the explicit (first row) and implicit (second
  row) IMEX SSP-schemes. We use the standard notation
  SSP$k(s,\sigma,p)$, where $k$ denotes the order of the SSP scheme
  and the triplet $(s,\sigma,p)$ characterizes respectively the number
  of stages of the implicit scheme ($s$), the number of stages of the
  explicit scheme ($\sigma$), and the order of the IMEX scheme ($p$).}
\begin{minipage}{1.6in}
{SSP2 $(3,3,2)$} 
\vskip 0.125cm
\begin{tabular} {c c c c c}
$ 0 $ & \vline & $0  $ & $0  $ & $0$ \\
$1/2$ & \vline & $1/2$ & $0  $ & $0$ \\
$ 1 $ & \vline & $1/2$ & $1/2$ & $0$ \\
\hline 
   & \vline & $1/3$ & $1/3$ & $1/3$ 
\end{tabular}
\end{minipage}
\vskip 0.2cm
\begin{minipage}{1.6in}
\begin{tabular} {c c c c c}
$1/4$ & \vline & $1/4$ & $0  $ & $0  $ \\
$1/4$ & \vline & $ 0 $ & $1/4$ & $0  $ \\
$ 1 $ & \vline & $1/3$ & $1/3$ & $1/3$ \\
\hline 
   & \vline & $1/3$ & $1/3$ & $1/3$ \\
\end{tabular}
\end{minipage}
\label{SSP2-322}
\vskip 0.5cm
\begin{minipage}{1.6in}
{SSP3 $(3,3,2)$}
\vskip .125cm
\begin{tabular} {c c c c c}
 $0  $ & \vline & $ 0 $ & $ 0 $ & $0$ \\
 $1  $ & \vline & $ 1 $ & $ 0 $ & $0$ \\
 $1/2$ & \vline & $1/4$ & $1/4$ & $0$ \\
\hline 
   & \vline & $1/6$ & $1/6$ & $2/3$ \\
\end{tabular}
\end{minipage}
\vskip 0.2cm
\begin{minipage}{1.6in}
\begin{tabular} {c c c c c}
$\gamma$   & \vline & $\gamma$       & $0$      & $0$   \\
$1-\gamma$ & \vline & $1-2 \gamma$   & $\gamma$ & $0$   \\
$ 1/2$     & \vline & $1/2 - \gamma$ & $0$      & $\gamma$ \\
\hline 
   & \vline & $1/6$ & $1/6$ & $2/3$ \\
\end{tabular}
\end{minipage}
\vskip 0.5cm
\begin{minipage}{1.6in}
%\centering
SSP3 $(4,3,3)$
\vskip 0.125cm
\begin{tabular} {c c c c c c}
 $0  $ & \vline & $0$  & $ 0 $ & $ 0 $ & $0$  \\
 $0  $ & \vline & $0$  & $ 0 $ & $ 0 $ & $0$  \\
 $1  $ & \vline & $0$  & $ 1 $ & $ 0 $ & $0$  \\
 $1/2$ & \vline & $0$  & $1/4$ & $1/4$ & $0$  \\
\hline 
   & \vline &  $0$ & $1/6$ & $1/6$ & $2/3$ \\
\end{tabular}
\end{minipage}
\vskip .2cm
\begin{minipage}{1.6in}
\centering
\begin{tabular} {c c c c c c}
 $\alpha$   & \vline & $\alpha$  &  $0$  &  $0$  & $0$  \\
 $0$        & \vline & $-\alpha$  &  $\alpha$  &  $0$  & $0$  \\
 $1$        & \vline & $0$  &  $1-\alpha$  &  $\alpha$  & $0$ \\
 $1/2$      & \vline & $\beta$  & $\eta$ & $1/2-\beta-\eta-\alpha$ & $\alpha$ \\
\hline 
   & \vline &  $0$ & $1/6$ & $1/6$ & $2/3$ \\
\end{tabular}
\end{minipage}
\begin{eqnarray}
 &\alpha \equiv 0.24169426078821\,, &\beta \equiv 0.06042356519705\,, \nonumber \\
 &\gamma \equiv 1 - {1}/{\sqrt{2}}\,, &\eta \equiv 0.12915286960590\,. \nonumber
\end{eqnarray}
\end{table}

%%%%%%%%%%%%%%%%%%%%%%%%%%%%%%%%%%%%%%%%%%%%
%%%%%%%%%%%%%%%%%%%%%%%%%%%%%%%%%%%%%%%%%%%%
%%%%%%%%%%%%%%%%%%%%%%%%%%%%%%%%%%%%%%%%%%%%

\section{IMEX Runge-Kutta scheme for the augmented resistive MHD equations}\label{section4}

Having reviewed the main properties of the IMEX schemes, we now apply
them to the particular case of the special relativistic resistive MHD
equations. Our goal is to consider a numerical implementation of the
general system that can deal with standard hydrodynamic issues (like
shocks and discontinuities) as well as those brought up by the stiff
terms discussed in the previous Section. Hence, we adopt high-resolution
shock-capturing algorithms (see Appendix~\ref{appendixB}) together
with IMEX schemes. Because the first ones involve the introduction of
conserved variables in order to cast the equations in a conservative
form, we first discuss how to implement the IMEX scheme within our
target system and subsequently how to perform the transformation 
from the conserved variables to the primitive ones.

%%%%%%%%%%%%%%%%%%%%%%%%%%%%%%%%%%%%%%%%%%%%%%%
\subsection{IMEX schemes for the Maxwell-Hydrodynamic equations and 
treatment of the implicit stiff part}

For our target system of equations it is possible to introduce a
natural decomposition of variables in terms of those whose evolution
do not involve stiff terms and those which do. More specifically, with
the electrical resistivity $\eta$ playing the role of the relaxation
parameter $\epsilon$, the vector of fields $\boldsymbol{U}$ can be
split in two subsets $\{\boldsymbol{X},\boldsymbol{Y}\}$, with
${\boldsymbol X}= \{ {\boldsymbol E} \}$ containing the stiff terms,
and ${\boldsymbol Y} = \{ {\boldsymbol B}, \psi, \phi, q, \tau,
{\boldsymbol S}, D \}$ the non-stiff ones.

Following the prototypical Eq.~(\ref{stiff_equation}), the evolution
equations for the relativistic resistive MHD equations can then be
schematically written as
\begin{eqnarray}\label{split}
    \partial_t {\boldsymbol Y} &=& F_{\!_{Y}}({\boldsymbol X},{\boldsymbol Y})\,,\\
    \partial_t {\boldsymbol X} &=& F_{\!_{X}}({\boldsymbol X},{\boldsymbol Y}) + \frac{1}{\epsilon({\boldsymbol Y})} R_{\!_{X}}({\boldsymbol X},{\boldsymbol Y})\,, 
\end{eqnarray}
where the relaxation parameter $\epsilon$ is allowed to depend also on
the ${\boldsymbol Y}$ non-stiff fields. The vector ${\boldsymbol Y}$
can be evolved straightforwardly as it involves no stiff term. We
further note that for our particular set of equations, it is
convenient to write the stiff part as
\begin{eqnarray}\label{stiff_part}
  R_{\!_{X}}({\boldsymbol X},{\boldsymbol Y}) = A({\boldsymbol Y})
  {\boldsymbol X} + S_{\!_{X}}({\boldsymbol Y})\,.
\end{eqnarray}  
As a result, the procedure to compute each stage
${\boldsymbol U}^{(i)}$
of the IMEX scheme can be performed in two steps:
\begin{enumerate}
  \item Compute the explicit intermediate values $\{ {\boldsymbol
    X}^{*}, {\boldsymbol Y}^{*} \}$ from all the
   previously known levels, that is
\begin{eqnarray}\label{first_stepa}
&& \hskip -.7cm
  {\boldsymbol Y}^{*} = {\boldsymbol Y}^n + \Delta t~
  \sum_{j=1}^{i-1}~ {\tilde{a}}_{ij} F_{\!_{Y}}({\boldsymbol U}^{(j)}) \,, 
 \\
\label{first_stepb}
&& \hskip -.7cm
  {\boldsymbol X}^{*} = {\boldsymbol X}^n + \Delta t~ \sum_{j=1}^{i-1}~ {\tilde{a}}_{ij} F_{\!_{X}}({\boldsymbol U}^{(j)}) 
     + \Delta t~ \sum_{j=1}^{i-1}~ \frac{a_{ij} }{\epsilon^{(j)}} R_{\!_{X}}({\boldsymbol U}^{(j)})  \,,
\nonumber \\
   \end{eqnarray} 
where we have defined $ \epsilon^{(j)} \equiv \epsilon({\boldsymbol
  Y}^{(j)})$ and $a_{ij}/\epsilon^{(j)}$ in Eq.~(\ref{first_stepb})
is a simple division and not a contraction on dummy indices.

   \item Compute the implicit part, which involves only ${\boldsymbol X}$, by solving
  \begin{eqnarray}
    \label{second_step_a}
   {\boldsymbol Y}^{(i)} &=& {\boldsymbol Y}^{*} \, , \\
    \label{second_step_b}
   {\boldsymbol X}^{(i)} &=& {\boldsymbol X}^{*} + \Delta t~ \frac{a_{ii} }{\epsilon^{(i)}} R_{\!_{X}}({\boldsymbol U}^{(i)}) \, .  
   \end{eqnarray} 
   Note that the implicit equation, with the previous assumption (\ref{stiff_part}),
   can be inverted  explicitly
   \begin{eqnarray}\label{invert_matrix}
       {\boldsymbol X}^{(i)} &=& M ({\boldsymbol Y}^*) ~
                    ( \boldsymbol{X}^* + a_{ii}~\frac{\Delta t}{\epsilon^{(i)}}~{\boldsymbol S}_{\!_{X}}({\boldsymbol Y}^*) ) \, , \\
       M({\boldsymbol Y}^*) &=& [I - a_{ii}~\frac{\Delta t} {\epsilon^{(i)}} A({\boldsymbol Y}^*)]^{-1} \,,
   \end{eqnarray}
since the form of the matrix $[I - a_{ii}~{\Delta t} A({\boldsymbol
    Y}^*)/{\epsilon^{(i)}} ]$ is known explicitly in terms of the
evolved fields.
\end{enumerate}

The explicit expressions for stiff part are then given simply by
\begin{eqnarray}\label{matrix1}
  {\boldsymbol R}_{\!_{E}} &=& -W {\boldsymbol E} + W~({\boldsymbol E} \cdot {\boldsymbol v}) {\boldsymbol v}
         - W {\boldsymbol v} \times {\boldsymbol B} \, , \\
  {\boldsymbol S}_{\!_{E}}  &=& - W {\boldsymbol v} \times {\boldsymbol B} \, ,
\end{eqnarray} 
with the matrix $A$ defined as
\begin{equation}
A \equiv W \left( \begin{array}{ccc}
 -1+v_x^2 & v_x~v_y & v_x~v_z \\
  v_x~v_y &-1+v_y^2 & v_y~v_z \\
  v_z~v_x & v_z~v_y &-1+v_z^2 \end{array} \right)\,.
\end{equation}
Hence, the matrix $M$ can be computed explicitly to obtain
\[ \frac{1}{m} \left( \begin{array}{ccc}
 a+W+a W^2 v_x^2 &\hskip-3mm a W^2 v_x v_y   & \hskip-3mm a W^2 v_x v_z \\
  a W^2 v_x v_y  &\hskip-3mm a+W+a W^2 v_y^2 & \hskip-3mm a W^2 v_y v_z \\
  a W^2 v_z v_x  &\hskip-3mm a W^2 v_z v_y   & \hskip-3mm a+W+a W^2 v_z^2  \end{array} \right)\]
where $m \equiv W^2 a+ W a^2 + W + a$ and $a \equiv a_{ii}~\sigma^{(i)}~\Delta t$.\\

Summarizing: First, an intermediate state $\{ \boldsymbol{E}^*\}$ is
found through the evolution of the non-stiff part for the electric
field. Second, if the velocity $ {\boldsymbol v}$ is known, the
evolution of the stiff part can be performed by acting with $M$ to obtain
\begin{eqnarray}\label{invert_matrix_E}
  {\boldsymbol E} = M ({\boldsymbol v}) ~
[ \boldsymbol{E^*} + a_{ii}~\Delta t~\sigma^{(i)}~{\boldsymbol S}_{\!_{E}}({\boldsymbol v},{\boldsymbol B}) ]\,.
\end{eqnarray}
At this point the approach proceeds with the conversion from the
conserved variables to the primitive ones. Because of the coupling
between the electric and the velocity fields, such a procedure is
rather involved and more complex than in the ideal-MHD case; a
detailed discussion of how to do this in practice will be presented in
Sect.~\ref{inversion_con2prim}.

It is interesting to highlight the consistency
at two known limits of the implicit solution of the stiff part. In the
ideal-MHD limit (\ie $\sigma \rightarrow \infty$) the first term of
Eq.~(\ref{invert_matrix_E}) vanishes, while the contribution of the
second term leads to the ideal-MHD condition (\ref{ef_imhd}). On the
other hand, in the vanishing conductivity limit (\ie $\sigma
\rightarrow 0$) the second term in Eq.~(\ref{invert_matrix_E})
vanishes, and the matrix reduces to the identity one $M({\boldsymbol
  v}) = I$. In this case, the electric field is obtained only by
evolving the explicit part, \ie ${\boldsymbol E} = \boldsymbol{E^*}$.

Finally, it is important to stress that one could, in principle, have considered
the alternative route of adopting instead 
${\boldsymbol X}= \{ {\boldsymbol E}, q\}$, so that the
right-hand-side of $q$ would be considered stiff with $R_q = 0$ and
$S_q = \nabla \cdot {\boldsymbol R}_{\!_{E}}$. However, this choice could lead
to spurious numerical oscillations in the solution since the fluxes of
$q$ can be discontinuous, while they would be evolved with an implicit
Runge-Kutta. As it has been shown under fairly general conditions,
high-order SSP schemes are necessarily explicit~\citep{GotShuTad:2001}, so
it follows that this part of the equations cannot be evolved with the implicit Runge-Kutta
unless a low-order scheme is implemented.

%%%%%%%%%%%%%%%%%%%%%%%%%%%%%%%%%%%%%%%%%%%%%%%
\subsection{Transformation of conserved variables to primitive ones}\label{inversion_con2prim}

As mentioned in the previous Section, in order to evolve our system of
equations, the fluxes $\{ \boldsymbol{F}_{\tau}, {\bf F}_{\boldsymbol
  S} ,\boldsymbol{F}_{D} \}$ must be computed at each timestep. These
fluxes depend on the primitive fields $\{ \rho, ~ p, ~ {\boldsymbol v}
, ~ {\boldsymbol E} , ~{\boldsymbol B}\}$, which must be recovered
from the evolved conserved fields $\{ D, ~ \tau , ~ {\boldsymbol S}, ~
{\boldsymbol E}, ~{\boldsymbol B} \}$. These quantities are related by
complicated equations which become transcendental except for
particularly simple equations of state (EOS). As a result, the
conversion must be in general pursued numerically and the primitive
variables are then given by the roots of the function
\begin{equation}\label{trascendental}
   f ( {\bar p}) = p(\rho,\epsilon) - {\bar p}\,,
\end{equation}
where $p(\rho,\epsilon)$ is given by the chosen EOS and ${\bar p}$ is
the trial value for the pressure eventually leading to the primitive
variables. 

Note that since ${\boldsymbol Y}^{(i)} = {\boldsymbol Y}^{*}$ [\cf
  Eq.~(\ref{second_step_a})], the values of the conserved quantities
$\{ D, ~ \tau , ~ {\boldsymbol S},~{\boldsymbol B} \}$ at time
$(n+1)\Delta t$ are obtained by evolving their non-stiff evolution
equations which, however, provide only an approximate solution for the
electric field $\{ \boldsymbol{E^*} \}$. As discussed in the previous
Section, the final solution for the electric field ${\boldsymbol E}$
requires the inversion of an implicit equation and, hence, is a
function of the velocity ${\boldsymbol v}$ and of the fields $\{
{\boldsymbol B}, \boldsymbol{E^*} \} $ [\cf
  Eq.~(\ref{invert_matrix_E})].  However, the velocity is a primitive
quantity and thus not known at the time $(n+1)\Delta t$. It is clear,
therefore, that it is necessary to obtain, at the same time, the
evolution of the stiff part of the equations and the conversion of the
conserved quantities into to the primitive ones. In what follows we
describe how to do this in practice using an iterative procedure.

\begin{enumerate}
  \item Adopt as initial guess for the velocity its value at the
    previous time level ${\boldsymbol v} = {\boldsymbol v}^n$.  The
    electric field ${\boldsymbol E}$ is computed by
    Eq.~(\ref{invert_matrix_E}) as a function of $({\boldsymbol
      E}^*,{\boldsymbol v},{\boldsymbol B})$.

   \item Adopt as initial guess for the pressure its value at the
     previous time level $p = p^{n}$. Compute in the
     following order

     \begin{eqnarray}\label{guesses}
       {\boldsymbol v} &=& \frac{{\boldsymbol S} - {\boldsymbol E} \times {\boldsymbol B}}
                      {\tau - (E^2 + B^2)/2 + p} \, , \nonumber \\
       W &=& \frac{1}{\sqrt{1 - v^2}}\,,  \nonumber \\
       \rho &=& \frac{D}{W} \, , \nonumber \\
       \epsilon &=& \frac{\tau - (E^2 + B^2)/2 - D~W + p~(1-W^2)}
                           {D~W} \, .
     \end{eqnarray}  
    \item Solve numerically Eq.~(\ref{trascendental}) by means of an
      iterative Newton-Raphson solver, so that the solution at the
      iteration $m+1$ can be computed as
     \begin{equation}\label{newton-raphson}
       p_{m+1} = p_m - \frac{f(p_m)}{f'(p_m)} \, .
     \end{equation}  
    The derivative of the function $f(p)$ needed for the Newton-Raphson solver
    can be computed as
    \begin{equation}\label{derivativef}
       f' (p) = v^2 c_{s}^2 - 1\,,
    \end{equation}
    with $c_{s}$ being the local speed of the fluid which, for an
    ideal-fluid EOS $p(\rho,\epsilon) = (\Gamma -1)~\rho~\epsilon$ is
    given by
    \begin{equation}
       \label{soundspeed}
       c_{s}^2 = \frac{\Gamma (\Gamma-1)  \epsilon}{ 1 + \Gamma~\epsilon}  \, .
    \end{equation}

    \item With the newly obtained values for the velocity
      ${\boldsymbol v}$ and the pressure $p$, the steps (i)--(iii) can
      be iterated until the difference between two successive values
      falls below a specified tolerance.
\end{enumerate}

The approach discussed above is a simple procedure that can be implemented
straightforwardly and works well for moderate ratios of
$|{\boldsymbol B}|^2/p$, converging in less than $10$ iterations both for
smooth electromagnetic fields and for discontinuous ones. 
Faster and more robust procedures to obtain the
primitive variables certainly ca be implemented, but this is beyond the
scope of this work.

%%%%%%%%%%%%%%%%%%%%%%%%%%%%%%%%%%%%%%%%%%%%%%%%%%%%%%%%%%%%%%%%%%%%%%%%%%%%%%%
%%%%%%%%%%%%%%%%%%%%%%%%%%%%%%%%%%%%%%%%%%%%%%%%%%%%%%%%%%%%%%%%%%%%%%%%%%%%%%%
%%%%%%%%%%%%%%%%%%%%%%%%%%%%%%%%%%%%%%%%%%%%%%%%%%%%%%%%%%%%%%%%%%%%%%%%%%%%%%%

\begin{figure}
\centerline{\includegraphics[width = 75mm]{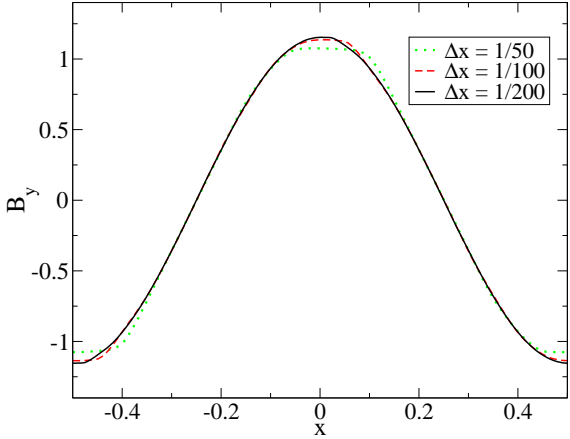}}
\caption{Magnetic field component $B_y$ for a large-amplitude CP
  Alfv\'en wave and for three different resolutions $\Delta
  x=\{1/50,1/100,1/200 \}$.  The conductivity is constant with a
  magnitude of $\sigma=10^6$. The agreement betweem
  the exact solution and that corresponding to the high resolution one
  is excellent.}
\label{alfven}
\end{figure}

\section{Numerical tests}\label{section5}

In this section we present several one-dimensional (1D) or
two-dimensional (2D) tests which have been used to validate the
implementation of the IMEX Runge-Kutta schemes in the different
regimes of relativistic resistive MHD.  In all these tests we employ
the ideal-fluid EOS with $\Gamma = 2$ for the 1D tests and $\Gamma =
4/3$ in the 2D ones. The different tests span several prescriptions
for the conductivity and compare the solutions obtained either with
those expected in the ideal-MHD limit or with those obtained with the
Strang-splitting technique.

More specifically, in 1D we consider large-amplitude circularly
polarized (CP) Alfv\'en waves to test the ability of the code to
reproduce the ideal-MHD results when adopting a very large
conductivity.  The intermediate conductivity regime is instead tested
by simulating a self-similar current sheet. Finally, a large range of
uniform and non-uniform conductivities are used for a representative
shock-tube problem.  In 2D, on the other hand, we first consider a
commonly employed test for ideal-MHD codes corresponding to a
cylindrical explosion. Subsequently, we simulate a toy model for a
``magnetized neutron star'' when modelled as a cylindrically symmetric
density distribution obeying a Gaussian-profile. The behaviour of the
magnetic field is studied again for a range of constant and
non-uniform conductivities.

\subsection{One-dimensional tests}

\begin{figure}
\centerline{\includegraphics[width = 75mm]{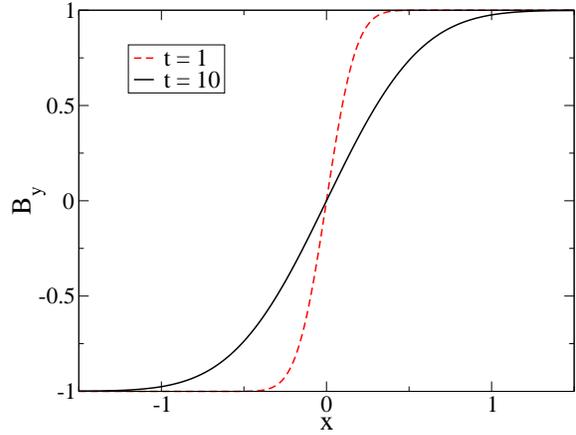}}
\caption{Magnetic field component $B_y$ in a self-similar current
  sheet. The solution is computed with $N=200$ gridpoints ($\Delta x = 1/200$) 
  and is shown
  at the initial time $t=1$ and at $t=10$. The conductivity is uniform
  with a magnitude of $\sigma=10^2$ (\ie $\eta = 1/\sigma=0.01$). The
  numerical solution is in excellent agreement with the
  exact one.}
\label{sheet}
\end{figure}

\subsubsection{Large amplitude CP Alfv\'en waves}

This test is discussed in detail by \citet{DelZanBucLon:2007} and we
report here only a short summary. The solution describes the
propagation of a large amplitude circularly-polarized Alfv\'en waves
along a uniform background field $B_0$ in a domain with periodic
boundary conditions. The exact solution in the ideal-MHD limit and
assuming $v_x = 0$ for simplicity, is given by
\citep{DelZanBucLon:2007}
\begin{eqnarray}
   (B_y,B_z) &=& \eta_A B_0 ~ (\cos[k(x-v_A~t)], \sin[k(x-v_A~t)])\,, \nonumber \\
   (v_y,v_z) &=& -\frac{v_A}{B_0} ~(B_y, B_z) \,,
\end{eqnarray}
where $B_x=B_0$, $k$ is the wave vector, $\eta_A$ is the amplitude
of the wave and the special relativistic Alfv\'en speed $v_A$ is
given by
\begin{equation}
   v_A^2 = \frac{2 B_0^2}{h + B_0^2 (1 + \eta_A^2)} \left(1 + \sqrt{1 - \left(\frac{2 \eta_A B_0^2}{h + B_0^2 (1 + \eta_A^2)} \right)^2 } \right)^{-1}\,.
\end{equation}
In practice, using such ideal-MHD solution it is possible to assess
the accuracy of evolution of the resistive equations by requiring that
for very large conductivities the numerical solution approaches the
exact one as the resolution is progressively increased. It is also
worth remarking that although we do not expect the solution of the
resistive MHD equations to converge to that of ideal MHD for any
finite value of $\sigma$, we also expect the differences between the
two to be ${\cal O}(v/\sigma)$ and thus negligibly small for
sufficiently large values. For this reason, we have performed the
evolution with a high uniform conductivity of $\sigma = 10^6$ for three
different resolutions $N=\{50,100,200\}$ covering the computational
domain $x \in [-0.5,0.5]$.  In addition, the initial data parameters
have been chosen so that $\rho = p = \eta_A = 1$ and $B_0 = 1.1547$,
thus yielding $v_A =1/2$, with a full period being achieved at $t=2$.

\begin{figure*}
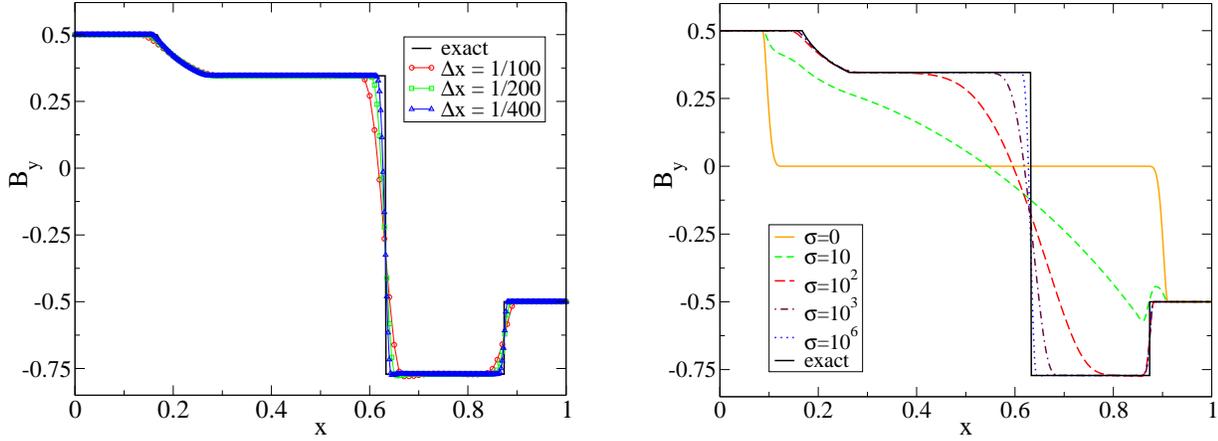

\centerline{
\includegraphics[width = 75mm]{shock_convergence.eps}
\hskip 1.0cm
\includegraphics[width = 75mm]{shock_condcte.eps}
}
\caption{\textit{Left panel:} Magnetic field component $B_y$ in the
  solution of the shock-tube problem. Different lines refer to three
  different resolutions and to the exact ideal-MHD solution at
  $t=0.4$.  The conductivity is uniform  with a magnitude of
  $\sigma_0=10^6$. \textit{Right panel:} The same as in the left panel
  but for different uniform conductivities. Note that for
  $\sigma_0=0$ the solution describes a discontinutiy 
  propagating at
  the speed of light and corresponding to Maxwell equations in
  vacuum. As the conductivity increases, the solution tends to the
  ideal-MHD one.}
\label{shock_convergence}
\end{figure*}

Fig.~\ref{alfven} confirms this expectation by reporting the component
$B_y$ after one period and thus overlapping with the initial one (at
$t=0$) for the highest resolution.  This test shows clearly that in
the limit of very high conductivity the resistive MHD equations tend
to a solution which is very close to the same solution obtained in the
ideal-MHD limit. The convergence rate measured for the different
fields is consistent with the second-order spatial discretization
being used as expected for smooth flows (see
Appendix~\ref{appendixB}).

\subsubsection{Self-similar current sheet}

\begin{figure}
\centerline{\includegraphics[width = 75mm]{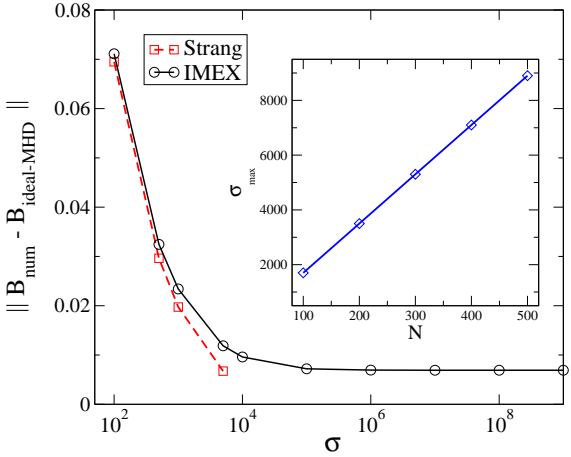}}
\caption{Differences in the magnetic field component $B_y$ between the
  numerical solution computed with either the Strang or the IMEX
  schemes and the exact solution of the shock-tube in the ideal-MHD
  limit. The differences are computed for several uniform
  conductivities, although the Strang-splitting technique does not
  yield a stable solution for values larger than $\sigma_0 \sim
  7000$ for the reference resolution of $\Delta x = 1/400$ (i.e. with
  $400$ gridpoints). Shown in the
  inset is the maximum conductivity for which a solution was possible,
  $\sigma_{\rm max}$, as a function of the number of gridpoints, $N$.}
\label{error}
\end{figure}

The details of this test are described by~\citet{Kom:2007}, so again
we provide here only a short description for completeness. We assume
that the magnetic pressure is much smaller than the fluid pressure
everywhere, with a magnetic field given by ${\boldsymbol B} = (0,
B_y(x,t), 0)$, where $B_y(x,t)$ changes sign within a thin current
layer of width $\Delta l$. Provided the initial solution is in
equilibrium ($p={\rm const.}$), the evolution is a slow diffusive
expansion of the layer due to the resistivity and described by the
diffusion equation [\cf Eq.~(\ref{newtonian_rMHD}) with ${\boldsymbol
    v}=0$]
\begin{eqnarray}\label{diffusion_eq}
    \partial_t B_y - \frac{1}{\sigma} \partial_x^2 B_y = 0 ~~.
\end{eqnarray}
As the system expands, the width of the layer becomes much larger
than $\Delta l$ and it evolves in a self-similar fashion. For $t > 0$, 
the analytical exact solution is given by
\begin{eqnarray}\label{diffusion_sol}
   B_y(x,t) = B_0 \, {\rm erf} \,\left(
   \frac{1}{2}\sqrt{\frac{\sigma}{\xi}} \right )\,,
\end{eqnarray}
where $\xi = t/x^2$ and ``${\rm erf}$'' is the error function. This
solution can be used for testing the moderate resistive
regime. Following~\citet{Kom:2007}, and in order to avoid the singular
behaviour at $t=0$, we have chosen as initial data the solution at
$t=1$ with $p=50$, $\rho=1$, $E={\boldsymbol v}=0$ and
$\sigma=100$. The domain covers the region $x \in [-1.5,1.5]$ with
$N=200$ points.

\begin{figure*}
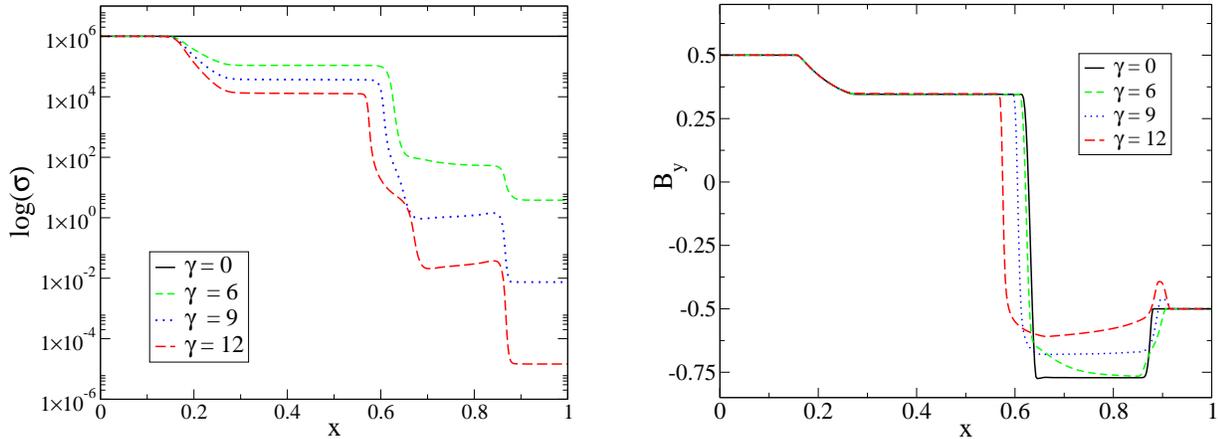

\centerline{
\includegraphics[width = 75mm]{shock_condpot.eps}
\hskip 1.0cm
\includegraphics[width = 75mm]{shock_pot.eps}
}
\caption{\textit{Left panel:} Evolution of a non-uniform conductivity
  $\sigma$ in the shock-tube problem for different values of $\gamma$
  and indicated by the different lines ($\sigma_0=10^6$ for all
  lines). Notice the large variability on the magnitude of the
  conductivity. \textit{Right panel:} The same as in the left panel
  but for the magnetic field component $B_y$.}
\label{shock_condpot}
\end{figure*}

The numerical simulation is evolved up to $t=10$ and then the
numerical and the exact solution are compared in Fig.~\ref{sheet}.
The two solutions match so well that they are not distinguishable on
the plot, thus, showing that the intermediate-conductivity regime is
also well described by our method.

\subsubsection{Shock-tube problem}

As prototypical shock-tube test we consider a simple MHD
version of the Brio and Wu test \citep{BriWu:1988}, where
the initial left and right states are separated at $x=0.5$ and are given by
\begin{eqnarray}
   (\rho^L,p^L,B_y^L) &=& (1.0, 1.0, 0.5)\,, \nonumber \\
   (\rho^R,p^R,B_y^R) &=& (0.125, 0.1, -0.5)\,, \nonumber
\end{eqnarray}
while all the other fields set to $0$. We consider both uniform and
non-uniform conductivities. In the latter case we adopt the following
prescription
\begin{equation}\label{def_cond}
   \sigma = \sigma_0 D^{\gamma} \, ,
\end{equation}
thus allowing for nonlinearities in the dependence of the conductivity
on the conserved quantity $D$. This is one of the simplest cases, but
in realistic situations a more general expression for the conductivity
can be assumed, where $\sigma$ is a function of both the rest-mass
density and of the specific internal energy, \ie $\sigma = \sigma
(\rho, \epsilon)$.

The exact solution of the ideal MHD Riemann problem was found
by~\citep{GiaRez:2006}, and in our particular case it has been
computed with a publicly available code
[see~\citet{GiaRez:2006}]. When $B_x=0$, the structure of the solution
contains only two fast waves, a rarefaction moving to the left and a
shock moving to the right, with a tangential discontinuity between
them. More demanding Riemann problems have also been performed but the
procedure to convert the conserved variables into the primitive ones
has shown in these case a lack of robustness for large ratios of
$|{\boldsymbol B}|^2/p$.

We have first considered the case of uniform ($\gamma=0$) and very
large conductivity ($\sigma_0=10^6$) as in this case we can use the
solution in the ideal-MHD limit as a useful guide. The profile of the
magnetic field component $B_y$ for three different resolutions $\Delta
x=\{1/100,1/200,1/400\}$ and the exact solution are shown in the left
panel of Fig.~\ref{shock_convergence} at $t=0.4$. Overall, the results
indicate that even in the presence of shocks our numerical solution of
the resistive MHD tends to the ideal-MHD solution as the resolution is
increased. It is also interesting to study the behaviour of the
solution for different values of the constant $\sigma_0$ while still
keeping a uniform conductivity (\ie $\gamma=0$). This is shown in the
right panel of Fig.\ref{shock_convergence}, which displays the
different solutions obtained, and where it is possible to see how they
change smoothly from a wave-like solution for $\sigma_0=0$ to the
ideal-MHD one for $\sigma_0=10^6$.

This set up is also useful to perform a comparison between the IMEX
and the Strang-splitting approaches. In Fig.~\ref{error} we show the
$L^1$-norm of the difference between the numerical solution obtained
with both schemes and the ideal-MHD exact solution, for different
values of the conductivity with $N=400$ points.

Several comments are in order. Firstly, the reported difference
between the numerical solution for the resistive MHD equations and the
ideal-MHD equations should not be interpreted as an error given that
the latter is not the correct solution of the equations. Hence, the
fact that the use of a Strang-splitting method yields smaller
differences is simply a measure of its ability of better capture steep
gradients. Secondly, while the IMEX approach does not show any sign of
instability for $\sigma_0$ ranging between $10^2$ and $10^9$, the
implementation adopting the Strang-splitting technique becomes
unstable for moderately high values of the conductivity and, at least
for the shock-tube problem, no numerical solution was possible for
$\sigma_0 \gtrsim 7000$ at the above resolution. Increasing the
resolution can help increase the maximum value of the resistivity
which can be handled, but since this gain is only linear with the
number of gridpoints aiming for higher conductivities results impractical. 
This is shown in
the inset of Fig.~\ref{error}, which reports the maximum conductivity
for which a solution was possible, $\sigma_{\rm max}$, as a function
of the number of gridpoints, $N$. Finally, we note that the difference
between the IMEX numerical solution and the exact ideal-MHD one
saturates between $\sigma_0 \sim 10^5-10^6$. This is not surprising
since the differences are expected to be ${\cal O}(1/\sigma)$, and
thus the saturation in the differences essentially provides a measure
of our truncation error at the resolution used.

A more challenging test is offered by the solution of the shock-tube
in the presence of a non-uniform conductivity. In particular, we have
considered the same initial states and the same non-uniform
conductivity discussed above, but used different values for the
exponent $\gamma$ in (\ref{def_cond}) while keeping $\sigma_0$
constant. The results of this test are shown in the left panel of
Fig.~\ref{shock_condpot}, where the conductivity is plotted at $t=0.4$
for several values of $\gamma$. Note that the conductivity traces the
evolution of the rest-mass density and that the solution can be found
also when $\sigma$ varies of almost $12$ orders of magnitude across
the grid. Similarly, the right panel of Fig.~\ref{shock_condpot}
displays the component $B_y$ for the different values of $\gamma$. It
should be stressed that because of the relation (\ref{def_cond})
between $\sigma$ and $\rho$, the region on the left has at this time a
very high conductivity and the numerical solution tends to the
ideal-MHD one. The opposite happens on the right region, where the
conductivity is lower for higher values of $\gamma$. Clearly, the
results presented in Fig.~\ref{shock_condpot} show that our
implementation can handle non-uniform (and quite steep) conductivity
profiles even in the presence of shocks.

\begin{figure*}
\centerline{
\includegraphics[width = 75mm]{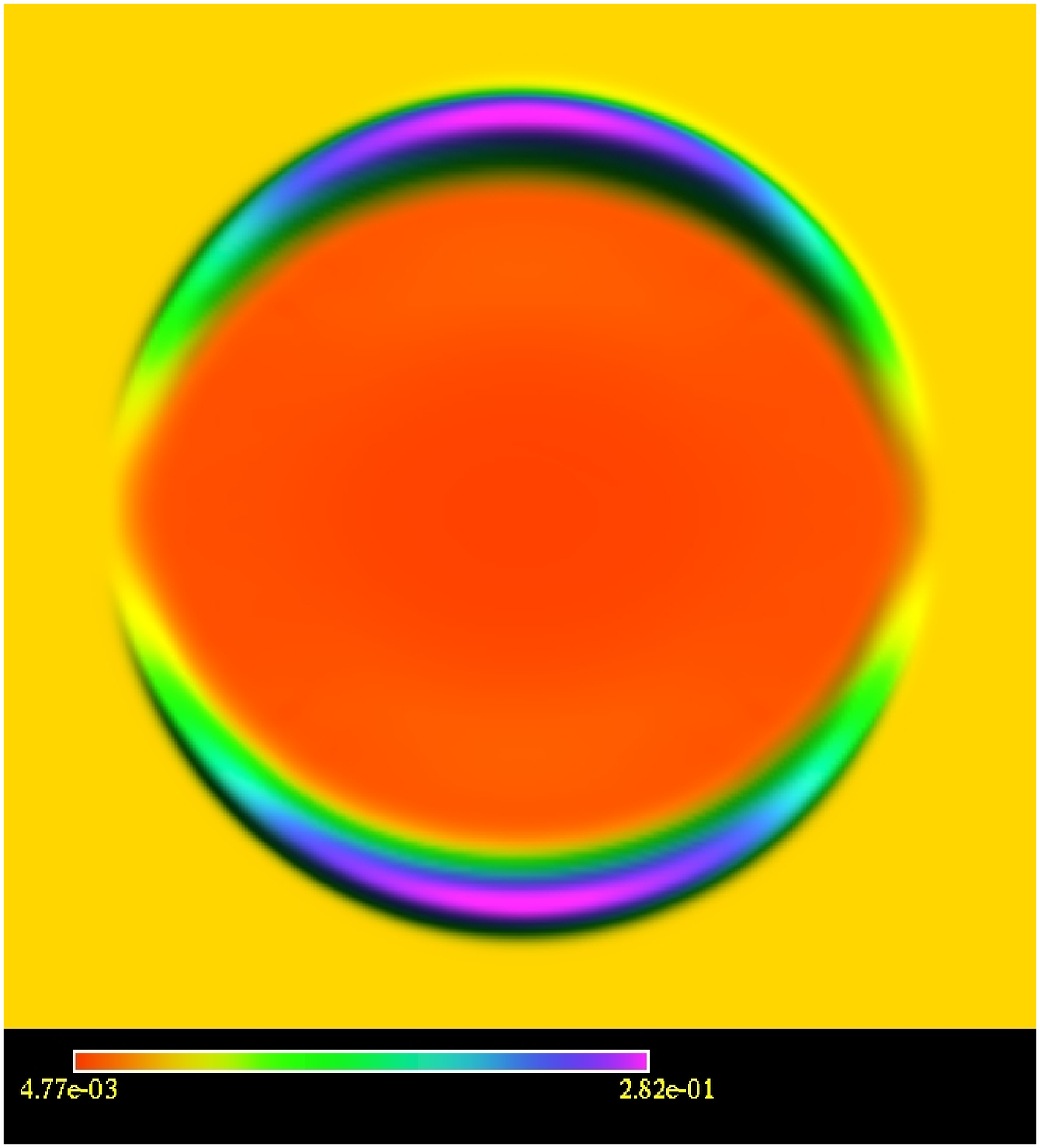}
\hskip 1.cm
\includegraphics[width = 75mm]{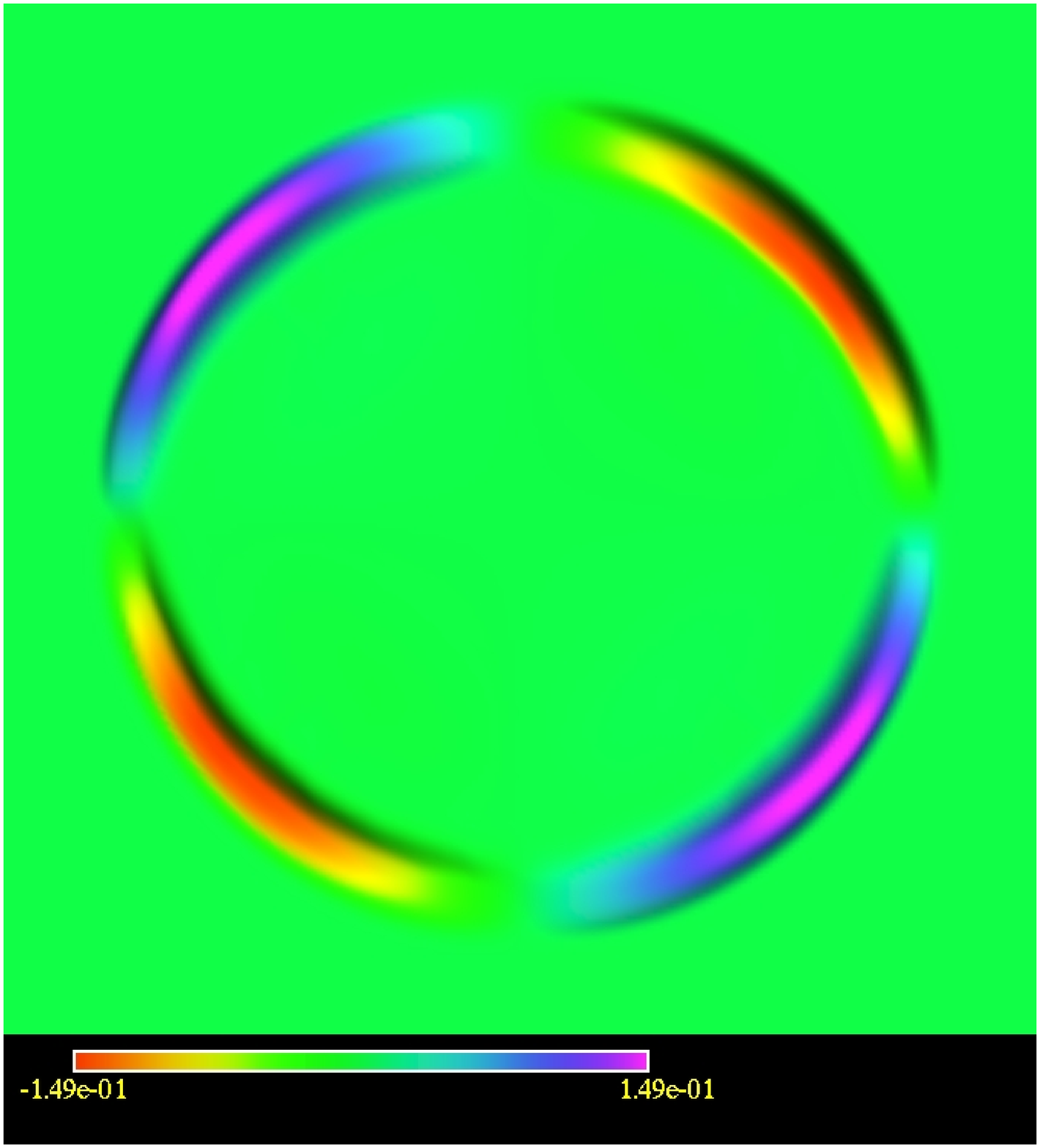} 
}
\caption{Magnetic field components $B_x$ (left panel) and $B_y$ (right
  panel) for the cylindrical explosion test at time $t=4$.}
\label{explosion_Bs_2D}
\end{figure*}

\subsection{Two-dimensional tests}

\subsubsection{The cylindrical explosion}

We now consider problems involving shocks in more than one dimension.
A demanding test for the relativistic codes is the cylindrical blast
wave expanding in a plasma with an initially uniform magnetic
field. Although there is no exact solution for this problem, strong
symmetric explosions are useful tests since shocks are present in all
the possible directions and the numerical implementation is therefore
tested in all of its parts.  For this test we set a square domain
$(x,y) \in [-6,6]$ with a resolution $\Delta x=\Delta y=1/200$. The
initial data is such that inside the radius $r<0.8$ the pressure is
set to $p=1$ while the density to $\rho=0.01$. In the intermediate
region $0.8 \leq r \leq 1.0$ the two quantities decrease exponentially
up to the exterior region $r>1$, where the ambient fluid has
$p=\rho=0.001$. The magnetic field is uniform with only one nontrivial
component ${\boldsymbol B}= (0.05, 0 ,0 )$. The other fields are set
to be zero (\ie ${\boldsymbol E} = q = 0$), which is consistent within
the ideal-MHD approximation.

The evolution is performed with a high conductivity $\sigma=10^6$ in
order to recover the solution from the ideal-MHD approximation. As
shown in Fig.~\ref{explosion_Bs_2D}, which reports the magnetic field
components $B_x$ (left panel) and $B_y$ (right panel) at time $t=4$,
we obtain results that are qualitatively similar to those published in
different
works~\citep{Kom:1999,NeiHirMil:2006,DelZanBucLon:2007,Kom:2007}. While
a strict comparison with an exact solution is not possible in this
case, the solution found matches extremely well the one obtained with
another 2D code solving the ideal MHD equations.
Most importantly, however, the figure
shows that the solution is regular everywhere and that similar results
can be obtained also with smaller values of the conductivity (\eg no
significant difference was seen for $\sigma \gtrsim 10^4$).

\subsubsection{The cylindrical star}

We next consider a toy model for a star, thought as an infinite column
of fluid aligned with the $z$-axis but with compact support in other
directions. Because of the symmetry in the $z$-direction, $\partial_z
\boldsymbol{U} = 0$ for all the fields and the problem is therefore
two-dimensional. More specifically, we consider initial data given by
\begin{eqnarray}
   \rho &=& \rho_o e^{-(r/r_o)^2} \, ,
\label{ID_fluid1} \\
   {\boldsymbol v} &=& (v^r,v^\phi,v^z) = \rho~( 0, \omega^\phi, 0) \, ,
\label{ID_fluid2} \\
   {\boldsymbol B} &=& (B^r,B^\phi,B^z) = \rho~\left( 0, 0, 2~B_o~(1 - \frac{r^2}{r_o^2})\right) \,,
\label{ID_fluid3} 
\end{eqnarray}
where $r \equiv \sqrt{x^2+y^2}$ is the cylindrical radial coordinate.
The other fields can be computed at the initial time by using the
polytropic EOS $p = \rho^\Gamma$, the ideal-MHD
expression~(\ref{ef_imhd}) for the electric field, and the electric
charge from the constraint equation $q = \nabla \cdot {\boldsymbol
  E}$.  We have chosen $r_o=0.7$, $\rho=1.0$, $\omega^\phi=0.1$ and
$B_0 = 0.05$. An atmosphere ambient fluid with $\rho=0.01$ is added
outside the cylinder. Finally, the resolution is $\Delta x = 1/200$
and the domain is $(x,y)\in [-3,3]$.

This simple problem exhibits some of the issues present in a
magnetized rotating neutron star: a compactly supported rest-mass
density distribution, an azimuthal velocity field and a poloidal
magnetic field.  Suitable source terms describing a gravitational
potential have been added to the Euler equations in order to get, at
least at the initial time, a stationary solution. In the ideal-MHD
limit the magnetic lines are frozen in the fluid and thus a static
profile is also expected for the magnetic field.

\begin{figure*}
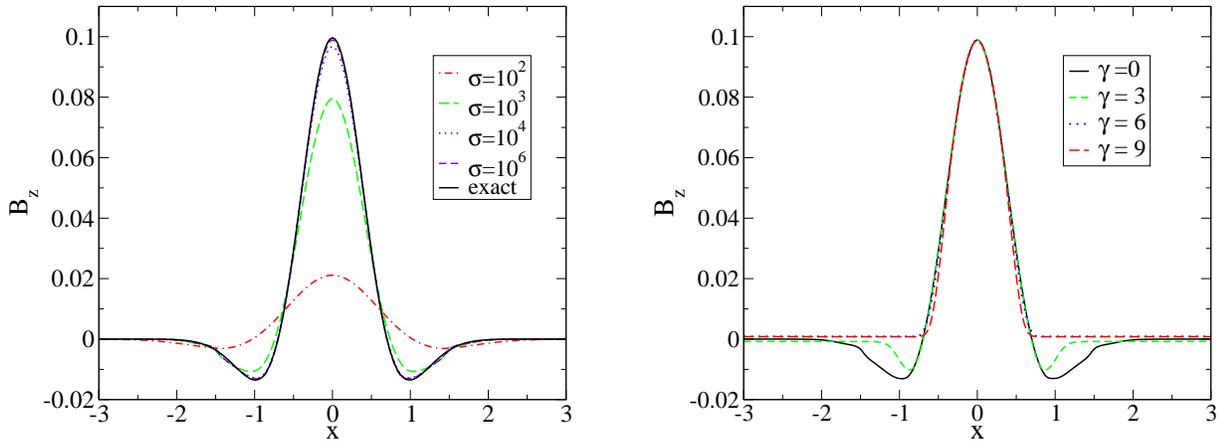

\centerline{
\includegraphics[width = 75mm]{star_Bz.eps}
\hskip 1.0cm
\includegraphics[width = 75mm]{star_Bp.eps}
}
\caption{\textit{Left panel:} Slice, at $y=0$, of the magnetic field
  component $B^z$ for different conductivities $\sigma$ and the exact
  solution in the ideal-MHD limit.  The resolution is $\Delta x=1/200$
  and the solution is plotted at $t=14$. \textit{Right panel:} the
  same configuration as in the left panel but with a non-uniform
  conductivity with $\sigma_0=10^6$ and $\gamma=[0,3,6,9]$. The values
  inside the star are essentially the same for any $\gamma$, while there
  are significant differences outside.}
\label{star_Bz}
\end{figure*}

In the left panel of Fig.~\ref{star_Bz} we plot the slice $y=0$ of the
magnetic field component $B^z$ at $t=14$ as obtained from the
evolution of the resistive MHD system for different uniform
conductivities in the range $\sigma_0~\epsilon~[10^2, 10^6]$. In the
limiting case $\sigma_0 = 0$ the solution corresponds to a wave
propagating at the speed of light (\ie the solution of the Maxwell
equations in vacuum), while for large values of $\sigma_0$ the
solution is stationary (as expected in the ideal-MHD limit). The
behaviour observed in the left panel Fig.~\ref{star_Bz} is also the
expected one: the higher the conductivity, the closer the solution is
to the stationary solution of the ideal-MHD limit. For low
conductivities, on the other hand, there is a significant diffusion of
the solution, which is quite rapid for $\sigma_0 < 10^2$ and for this
reason those values are not plotted here. We note that values of the
conductivity larger than $\sigma_0 > 10^7$ lead to numerical
instabilities that we believe are coming from inaccuracies in the
evolution of the charge density $q$, and which contains spatial
derivatives of the current vector.  In addition, the stiff quantity
$E_x$ is seen to converge only to an order $ \sim 1.5$. This can be
due to the ``final layer'' problem of the IMEX methods, which is known
to produce a degradation on the accuracy of the stiff
quantities. Luckily, this does not spoil the convergence of the
non-stiff fields, which are instead second-order convergent. It is
possible that the use of stiffly-accurate schemes can solve this
degradation of the convergence and this is an issue we are presently
exploring.

We finally consider the same test, but now employing the non-uniform
conductivity given by Eq.~(\ref{def_cond}) with $\sigma_0=10^6$ and
different values for $\gamma$. The results are presented in the right
panel of Fig.~\ref{star_Bz}, which shows that the magnetic fields
inside the star are basically the same in all the cases, stressing the
fact that the interior of the star will not be significantly affected
by the exterior solution, which has much smaller
conductivity. However, the electromagnetic fields outside the star do
change significantly for different values of $\gamma$, underlining the
importance of a proper treatment of the resistive effects in those
regions of the plasma where the ideal-MHD approximation is not a good
one.

%%%%%%%%%%%%%%%%%%%%%%%%%%%%%%%%%%%%%%%%%%%%%%%
\section{Conclusions}
\label{section6}

We have introduced Implicit-Explicit Runge-Kutta schemes to solve
numerically the (special) relativistic resistive MHD equations and
thus deal, in an effective and robust way, with the problems inherent
to the evolution of stiff hyperbolic equations with relaxation terms.
Since for these methods the only limitation on the size of the
timestep is set by the standard CFL condition, the approach suggested
here allows to solve the full system of resistive MHD equations
efficiently without resorting to the commonly adopted limit of the
ideal-MHD approximation.

More specifically, we have shown that it is possible to split the
system of relativistic resistive MHD equations into a set of equations
that involves only non-stiff terms, which can be evolved
straightforwardly, and a set involving stiff terms, which can also be
solved explicitly because of the simple form of the stiff
terms. Overall, the only major difficulty we have encountered in
solving the resistive MHD equations with IMEX methods arises in the
conversion from the conserved variables to the primitive ones. In this
case, in fact, there is an extra difficulty given by the fact that
there are four primitive fields which are unknown and have to be
inverted simultaneously. We have solved this problem by using extra
iterations in our 1D Newton-Raphson solver, but a multidimensional
solver is necessary for a more robust and efficient implementation of the
inversion process.

With this numerical implementation we have carried out a number of
numerical tests aimed at assessing the robustness and accuracy of the
approach, also when compared to other equivalents ones, such as the
Strang-splitting method recently proposed by~\citet{Kom:2007}. All of
the tests performed have shown the effectiveness of our approach in
solving the relativistic resistive MHD equations in situations
involving both small and large uniform conductivities, as well as
conductivities that are allowed to vary nonlinearly across the
plasma. Furthermore, when compared with the Strang-splitting
technique, the IMEX approach has not shown any of the instability
problems that affect the Strang-splitting approach for flows with
discontinuities and large conductivities.

While the results presented here open promising perspectives for the
implementation of IMEX schemes in the modelling of relativistic
compact objects, at least two further improvements can be made with
minor efforts. The first one consists of the generalization of the
(special) relativistic resistive MHD equations with a scalar isotropic
Ohm's law to the general relativistic case, and its application to
compact astrophysical bodies such a magnetized binary neutron
stars~\citep{AHLLMNPT:2008,LiuShaZacTan:2008}. The solution of the
resistive MHD equations can yield different results not only in the
dynamics of the magnetosphere produced after the merger, but also
provide the possibility to predict, at least in some approximation,
the electromagnetic radiation produced by the merger of these
objects. The second improvement consists of considering a non-scalar
and anisotropic Ohm's law, so that the behaviour of the currents in
the magnetosphere can be described by using a very high conductivity
along the magnetic lines and a negligibly small one in the transverse
directions~\citep{Kom:2004}. Such an improvement may serve as a first
step towards an alternative modelling of force-free plasmas.

%%%%%%%%%%%%%%%%%%%%%%%%%%%%%%%%%%%%%%%%%%%%%%%%%%%%%%%%%%%%%%
\section*{Acknowledgments}
%%%%%%%%%%%%%%%%%%%%%%%%%%%%%%%%%%%%%%%%%%%%%%%%%%%%%%%%%%%%%%
We would like to thank Eric Hirschmann, Serguei Komissarov, Steve
Liebling, Jonathan McKinney, David Neilsen and Olindo Zanotti for
useful comments and Bruno Giacomazzo for comments and for providing
the code computing the exact solution of the Riemann problem in ideal
MHD. LL and CP would like to thank FaMAF (UNC) for hospitality. CP is
also grateful to Lorenzo Pareschi for the many clarifications about
the IMEX schemes. This work was supported in part by NSF grants
PHY-0326311, PHY-0653369 and PHY-0653375 to Louisiana State
University, the DFG grant SFB/Transregio~7, CONICET and Secyt-UNC.

%%%%%%%%%%%%%%%%%%%%%%%%%%%%%%%%%%%%%%%%%%%%%%%
%%%%%%%%%%%%%%%%%%%%%%%%%%%%%%%%%%%%%%%%%%%%%%%
%%%%%%%%%%%%%%%%%%%%%%%%%%%%%%%%%%%%%%%%%%%%%%%
%%%%%%%%%%%%%%%%%%%%%%%%%%%%%%%%%%%%%%%%%%%%%%%

\appendix

%%%%%%%%%%%%%%%%%%%%%%%%%%%%%%%%%%%%%%%%%%%%%%%

\section{TVD space discretization}\label{appendixB}

We are generically interested in solving hyperbolic conservation laws
of the form
\begin{eqnarray}\label{hyperb_law}
   \partial_t \boldsymbol{U} + \partial_k~ ^k{\boldsymbol
     F}(\boldsymbol{U}) = {\boldsymbol S}(\boldsymbol{U}) \,,
\end{eqnarray}
where $\boldsymbol{U}$ is the vector of the evolved fields,
$^k{\boldsymbol F}$ are their fluxes and ${\boldsymbol S}$ contains
the sources terms. The semi-discrete version of this equation,
in one dimension, is simply given by 
\begin{eqnarray}\label{hyperb_law_semidiscrete}
   \partial_t \boldsymbol{U}_i = - \frac{\boldsymbol{\hat{F}}_{i+1/2} - \boldsymbol{\hat{F}}_{i-1/2}}
   {\Delta x} + {\boldsymbol S}(\boldsymbol{U}_i)\,,
\end{eqnarray}
where $\boldsymbol{\hat{F}}_{i \pm 1/2}$ are consistent numerical
fluxes evaluated at the interfaces between numerical cells. These
consistent fluxes are computed by using HRSC methods, which are based
on the use of Riemann solvers. More specifically, we have implemented
a modification of the Local Lax-Friedrichs approximate Riemann solver
introduced by~\citet{ABBM:2007}, which only needs the spectral radius
(\ie the maximum eigenvalue) of the system. In highly relativistic
cases, like the ones we are interested in, the spectral radius is
close to the light speed $c=1$ and so the Local Lax-Friedrichs reduces
to the simpler Lax-Friedrichs flux
\begin{eqnarray}\label{Lax-Friedrichs}
   {\hat{F}}_{i+1/2} = \frac{1}{2} [F_L + F_R + (u_L - u_R)]\,,
\end{eqnarray}
where $u_L,u_R$ are the reconstructed solutions on the left and on the
right of the interface and $F_L,F_R$ their corresponding fluxes. The
standard procedure is then to reconstruct the solution $u_L,u_R$ by
interpolating with a polynomial and then compute the fluxes
$F_L=F(u_L)$ and $F_R=F(u_R)$.  In our implementation we first
recombine the fluxes and the solution as~\citep{ABBM:2007} 
\begin{eqnarray}\label{recomb_fluxes1}
   {F}_{i}^{\pm} = {F}_{i} \pm u_i\,.
\end{eqnarray}
Then, using a piecewise linear reconstruction, these combinations can
be computed on the left/right of the interface as
\begin{equation}\label{recomb_fluxes2}
   {F}_{L}^{+} = {F}_{i}^{+}  + \frac{1}{2} \Delta_i^+  \,,
   \hskip 1.0cm
   {F}_{R}^{-} = {F}_{i+1}^{-}  - \frac{1}{2} \Delta_{i+1}^- \,,
\end{equation}
where $\Delta_i^{\pm}$ are just the slopes used to extrapolate
$F_i^{\pm}$ to the interfaces.  Finally, the consistent flux is
computed by a simple average
\begin{eqnarray}\label{Lax-Friedrichs2}
   {\hat{F}}_{i+1/2} = \frac{1}{2} [F_L^+ + F_R^-]\,.
\end{eqnarray}
For a linear reconstruction the slopes can be written as
\begin{eqnarray}\label{slope_fluxes}
   {\Delta}_{i}^{+} &=& L ( {F}_{i+1}^{+} - {F}_{i}^{+} ,
                          {F}_{i}^{+} - {F}_{i-1}^{+} )\,,   \nonumber \\
   {\Delta}_{i+1}^{-} &=& L ( {F}_{i+2}^{+} - {F}_{i+1}^{+} ,
                          {F}_{i+1}^{+} - {F}_{i}^{+} )\,,
\end{eqnarray}
so that it is trivial to check that the standard Lax-Friedrichs
(\ref{Lax-Friedrichs}) is recovered when $\Delta_i^+ = \Delta_i^-$.
The choice of these slopes becomes crucial in the presence of shocks
or very sharp profiles, while the use of some nonlinear operators
$L(x,y)$ preserves the Total Variation Diminishing (TVD) condition on
the interpolating polynomial. In this way, the TVD schemes capture
accurately the dynamics of strong shocks without the oscillations
which appear with standard finite-difference
discretizations. Monotonicity is typically enforced by making use of
slope limiters and we have in particular implemented the Monotonized
Centered (MC) limiter
\begin{equation}\label{limiter_MC}
   L(x,y) = \frac{1}{2} [{\rm sign}(x) + {\rm sign}(y)] ~{\rm
     min}(2|x|,2|y|,\frac{1}{2}|x+y|) \,,
\end{equation}
which provides a good compromise between robustness and accuracy. Note
that with linear reconstruction the scheme is second-order accurate in
the smooth regions, although it drops to first order near shocks and
at local extrema.

%%%%%%%%%%%%%%%%%%%%%%%%%%%%%%%%%%%%%%%%%%%%%%%%%%%%%%%%%%%%%%%%%%%%
%
%   B I B L I O G R A P H Y
%
%%%%%%%%%%%%%%%%%%%%%%%%%%%%%%%%%%%%%%%%%%%%%%%%%%%%%%%%%%%%%%%%%%%%

\bibliographystyle{mn2e}
%\bibliography{biblio}

\label{lastpage}

\end{document}